\documentclass[conference]{IEEEtran}

\usepackage{graphicx}
\usepackage{amsmath}
\usepackage{amssymb}
\usepackage{url}
\usepackage{multirow}
\usepackage{subcaption}
\usepackage{xcolor}
\usepackage{xspace}
\usepackage{nicefrac}
\usepackage{numprint}
\usepackage{algorithm}
\usepackage[noend]{algpseudocode}

\newcommand{\ournameNoSpace}{\emph{\mbox{DEMASQ}}}
\newcommand{\ourname}{\ournameNoSpace\xspace}

\DeclareMathAlphabet{\mathcal}{OMS}{cmsy}{m}{n}

\DeclareCaptionLabelFormat{custom}
{
      #1 \textbf{(#2)}
}

\DeclareCaptionFormat{custom}
{
    \small {#1#2 #3}
}

\makeatletter 
\newcommand{\linebreakand}{
  \end{@IEEEauthorhalign}
  \hfill\mbox{}\par
  \mbox{}\hfill\begin{@IEEEauthorhalign}
}
\makeatother

\usepackage{tikz}
\usepackage{amsmath}
\usepackage{amsmath}
\usepackage{enumitem}

\DeclareMathOperator*{\argmin}{arg\,min}
\makeatletter
\def\@xfootnote[#1]{
  \protected@xdef\@thefnmark{#1}
  \@footnotemark\@footnotetext}
\makeatother
\usepackage{footnote}

\begin{document}
\title{DEMASQ: Unmasking the ChatGPT Wordsmith}

\author{\IEEEauthorblockN{Kavita Kumari}
\IEEEauthorblockA{Technical University of Darmstadt\\ kavita.kumari@trust.tu-darmstadt.de}
\and
\IEEEauthorblockN{Alessandro Pegoraro}
\IEEEauthorblockA{Technical University of Darmstadt\\ alessandro.pegoraro@trust.tu-darmstadt.de}
\and
\IEEEauthorblockN{Hossein Fereidooni}
\IEEEauthorblockA{Technical University of Darmstadt\\ hossein.fereidooni@trust.tu-darmstadt.de}
\and
\linebreakand
\IEEEauthorblockN{Ahmad-Reza Sadeghi}
\IEEEauthorblockA{Technical University of Darmstadt\\ ahmad.sadeghi@trust.tu-darmstadt.de}
}

\IEEEoverridecommandlockouts
\makeatletter\def\@IEEEpubidpullup{6.5\baselineskip}\makeatother
\IEEEpubid{\parbox{\columnwidth}{
    Network and Distributed System Security (NDSS) Symposium 2024\\
    26 February - 1 March 2024, San Diego, CA, USA\\
    ISBN 1-891562-93-2\\
    https://dx.doi.org/10.14722/ndss.2024.231190\\
    www.ndss-symposium.org
}
\hspace{\columnsep}\makebox[\columnwidth]{}}

\maketitle

\begin{abstract}
The potential misuse of ChatGPT and other Large Language Models (LLMs) has raised concerns regarding the dissemination of false information, plagiarism, academic dishonesty, and fraudulent activities. Consequently, distinguishing between AI-generated and human-generated content has emerged as an intriguing research topic. However, current text detection methods lack precision and are often restricted to specific tasks or domains, making them inadequate for identifying content generated by ChatGPT.

\noindent In this paper, we propose an effective ChatGPT detector named \ourname, which accurately identifies ChatGPT-generated content. Our method addresses two critical factors: (i) the distinct biases in text composition observed in human- and machine-generated content and (ii) the alterations made by humans to evade previous detection methods. \ourname is an energy-based detection model that incorporates novel aspects, such as (i) optimization inspired by the Doppler effect to capture the interdependence between input text embeddings and output labels, and (ii) the use of explainable AI techniques to generate diverse perturbations. 

To evaluate our detector, we create a benchmark dataset comprising a mixture of prompts from both ChatGPT and humans, encompassing domains such as medical, open Q\&A, finance, wiki, and Reddit. Our evaluation demonstrates that \ourname achieves high accuracy in identifying content generated by ChatGPT.
\end{abstract}

\section{Introduction}
\label{sec:introduction}
The release of ChatGPT, an AI chatbot developed by OpenAI, has generated significant interest and extensive discussions within the Natural Language Processing (NLP) community and various other domains. ChatGPT, which was launched in November 2022, harnesses the capabilities of OpenAI's language models from the GPT-3.5 and GPT-4 series. By employing supervised and reinforcement learning techniques based on Human Feedback \cite{christiano2017deep, lambert2022illustrating}, ChatGPT generates sophisticated responses to diverse queries across various NLP fields, as highlighted in recent studies~\cite{biswas2023chatgpt, dowling2023chatgpt, omar2023chatgpt, susnjak2023applying, yeo2023assessing}.

The media's promotion of ChatGPT has generated various responses. News and media companies are employing it to enhance content creation, educators and academics are utilizing it to support course objectives, and individuals are taking advantage of its language translation capabilities. Unfortunately, advanced technologies like ChatGPT often face instances of misuse. Students are using it to generate assignments and coding projects \cite{bleumink2023keeping, cotton2023chatting}, scholars rely on it for producing academic papers \cite{gao2022comparing}, and malicious actors exploit it to spread fake news on social media platforms \cite{hacker2023regulating, de2023chatgpt}. Furthermore, ChatGPT has the potential to create seemingly realistic stories that can deceive unsuspecting readers \cite{bang2023multitask, shen2023chatgpt}. Therefore, the development of an effective detection algorithm capable of distinguishing AI-generated text, particularly from ChatGPT, from human-generated text has become a significant focus for researchers. 

When detecting AI-generated text using machine learning, two common approaches are black-box and white-box detection. Black-box detection involves accessing language models through APIs, posing challenges in identifying synthetic texts effectively~\cite{tang2023science}. It includes tasks like data collection, feature extraction, and constructing specialized classifiers, often using simpler ones like binary logistic regression~\cite{solaiman2019release}. In contrast, white-box detection has unrestricted access to language models, providing greater control over behavior and traceable results~\cite{tang2023science}. Zero-shot detection is an example of white-box detection, using pre-trained generative models like GPT-2 or Grover, along with fine-tuned language models for the detection task~\cite{mitchell2023detectgpt, solaiman2019release, zellers2019defending}.

The development of text detectors has attracted substantial attention in recent research, with considerable efforts devoted to developing detectors capable of distinguishing text generated by AI bots~\cite{cotton2023chatting, gao2022comparing, gehrmann2019gltr, perplexityAnalysis, khalil2023will, kumarage2023stylometric, kushnareva2021artificial, mitchell2023detectgpt, GPTZero, solaiman2019release, zellers2019defending, AITextDetector}. Moreover, certain assertions have been made concerning the discriminative ability of AI-text detectors in distinguishing ChatGPT-generated text from human-generated text~\cite{originality, aicontentdetector, bleumink2023keeping, copyleaks, DAG, Huggingface, Frohling2021feature, guo2023close, mitrovic2023chatgpt, AITextClassifier, writefull, WriterAIContentDetector}.

Our paper's primary focus revolves around the development of detectors to effectively distinguish between text generated by ChatGPT and humans. In our pursuit of this goal, we conducted a comprehensive evaluation of existing schemes and tools dedicated to the detection of ChatGPT-generated content. Our evaluation of the existing ChatGPT detectors has revealed that, in the best-case scenario, these methods achieved an accuracy rate of 47\%, as elaborated in Table \ref{tab:summary}.

\noindent \textbf{Our goals and contributions.}
We address the limitations of existing ChatGPT detectors and present our novel approach, inspired by the intricate nature of human communication. Human interactions exhibit diverse patterns and inherent biases, influenced by personal backgrounds, environmental factors, and the unique individuals involved. In contrast, machines interact based on predefined rules, maintaining consistent energy and frequency regardless of the human query. To account for this inherent bias, we leverage the Doppler effect to model the communication dynamics between humans and machines. This novel approach allows us to quantify the responsiveness of both humans and machines to questions posed by humans, thus providing a unique perspective on bias detection.

We extend the Doppler effect's application to text, hypothesizing that thoughts translated into text follow a similar physical process as phonetic energy waves in human communication. And by adopting drumhead vibrations we model waves and source frequency. Since the energy level is directly linked to its higher or lower frequency, we leverage this relationship and choose to develop and train an Energy-Based Model (EBM). In EBMs, energy serves as a key metric for understanding the connection between the model's input and output, with lower energy configurations representing the most optimal relationship between these variables. This model facilitates the discrimination between responses generated by humans and ChatGPT through analyzing their respective energy levels, effectively capturing the biases introduced by both entities. 

We stress that, no existing ChatGPT detector has explored these aspects, giving our novel design, named \ourname, a significant performance advantage as it achieves an high  
accuracy of 97\% on a representative benchmark dataset that contains diverse prompts from both ChatGPT and humans, spanning different domains which we detail in Section \ref{sub:dataset}.

Moreover, we address ChatGPT's rephrasing techniques by incorporating the explainable AI concept of Integrated Gradients (IG) \cite{sundararajan2017axiomatic} to perturb the input string's embedding. This perturbation allows us to extract crucial features, compute their energies, and identify a natural threshold between ChatGPT and human-like responses during the string's evolution. This integration of energy-based modeling, IG perturbation, and optimization enhances our detector's ability to differentiate between human and ChatGPT-generated responses while accounting for biases and rephrasing techniques.

In summary, our contributions in this paper include:
\begin{itemize} [noitemsep]
\item The design and implementation of \ourname, a novel ChatGPT detector for accurate discrimination between ChatGPT and human-generated text.
\item A unique approach that incorporates the effect of bias introduced during human interactions and rephrasing techniques that modify ChatGPT-generated responses to evade detection. We utilizing frequency and energy modeling, and explainable AI (cf. Section~\ref{sec:design}).
\item  The creation of a comprehensive benchmark dataset comprising over 100,000 samples and spanning various domains, such as wiki, reddit, openqa, finance, medical, arxiv, and political, and the rigorous evaluation of our approach's robustness across diverse contexts (cf. Section~\ref{sec:eval}). We evaluated the existing ChatGPT detectors on this benchmark, revealing that these methods achieve an accuracy rate of 47\% in the best-case scenario (Table \ref{tab:summary}). In contrast, \ourname achieves an accuracy of 96.5\% on the above mentioned combined benchmark dataset, significantly outperforming the existing detectors.
\item Our research sheds light on the current state of detection techniques and aims to stimulate further exploration in this critical field.
\end{itemize}

While \ourname demonstrates a significant superiority over the existing ChatGPT detectors, we acknowledge that there are still several challenges to be tackled. Firstly, we expect that \ourname, given its generic design, will exhibit comparable performance when subjected to prompts generated by other generative language models. Hence, we are currently investigating this aspect as part of our ongoing research efforts. Secondly, although we have demonstrated the superior performance of \ourname when compared to recent work on hybrid texts (as discussed in Section \ref{sub:rephrase}), we are actively refining our methodology to bolster its effectiveness in distinguishing content generated by ChatGPT, especially when confronted with rephrased or adversarial prompts.
\section{Background}
\label{sec:background}
In this section, we present a concise overview of the essential background knowledge encompassing all the related concepts necessary to comprehend the design of our proposed detection method. By providing this contextual information, we aim to equip readers with the foundational understanding required to grasp the intricacies and rationale behind our innovative detection approach.

\subsection{Large Language Models (LLMs)}
\label{sub:llm}
Large Language Models (LLMs) are advanced AI systems designed to understand and generate human-like text. They undergo extensive training on diverse datasets from various sources like books, articles, and websites. This training enables LLMs to produce coherent and contextually relevant responses. Examples of notable LLMs include GPT-3, GPT-2, Transformer-XL, BERT, and ChatGPT. These models represent just a fraction of the diverse range of LLMs developed by different organizations and researchers, each with its own strengths and areas of specialization.

\subsection{Doppler Effect}
\label{sub:dopp}
The Doppler effect occurs when the relative motion between a wave source and an observer exists. It manifests as a change in frequency of source as perceived by the observer. When a wave source moves toward an observer, there is an apparent increase in frequency, while the frequency decreases when the source moves away from the observer. This relationship between the observed frequency, denoted as $E_{f}$, and the emitted frequency, denoted as $E_{f_{\text{0}}}$, can be described by the following equation:

\begin{equation}
    E_{f} = \left( \frac{c_{v} \pm v_{r}}{c_{v} \pm v_{s}} \right)E_{f_{\text{0}}}  \label{eq:doppler}
\end{equation}

\begin{figure}[t]
\centering
\includegraphics[width=\linewidth]{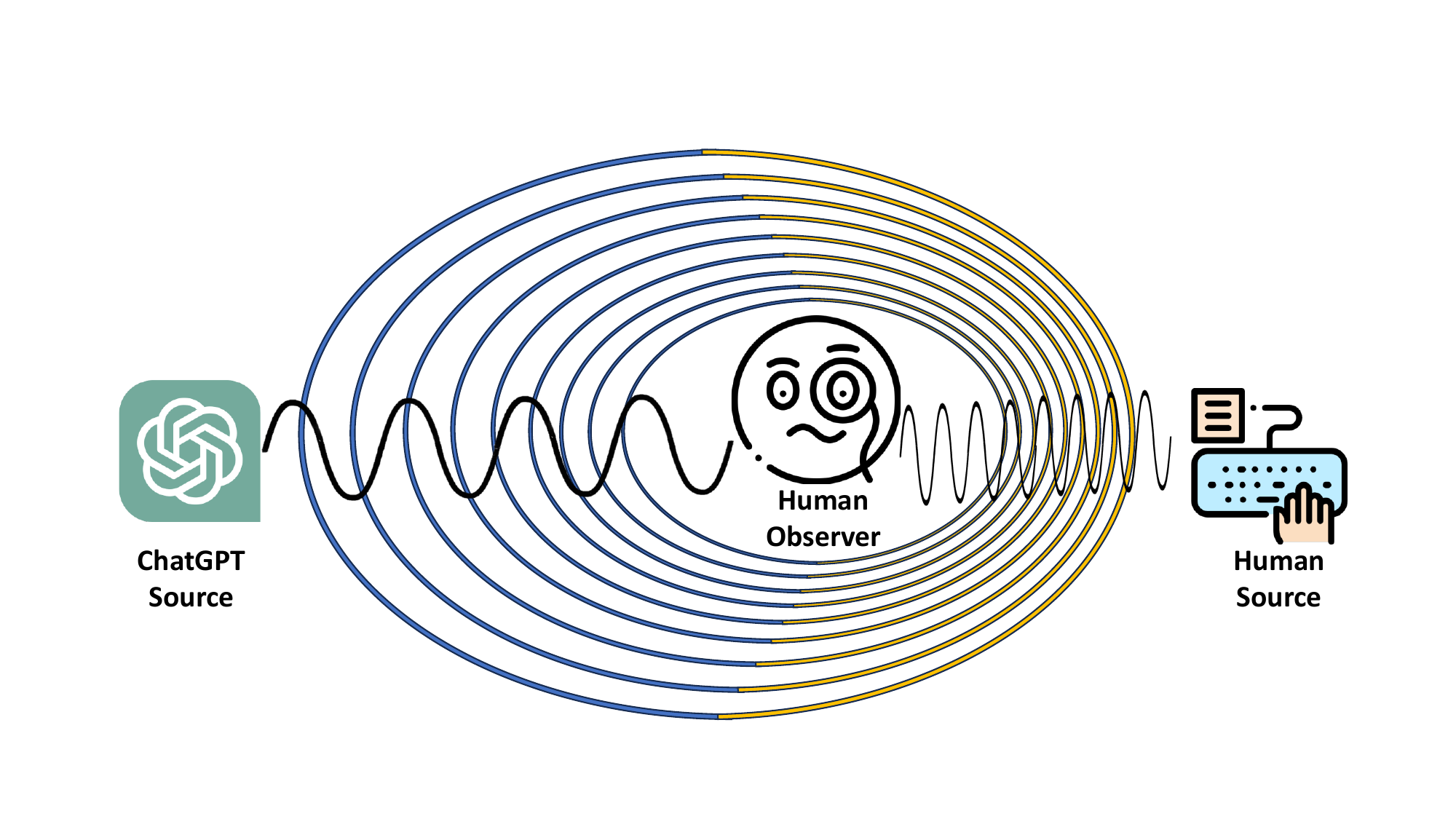}
\caption{Doppler effect.}
\label{fig:do_ef}
\end{figure}

In the above equation, $c_{v}$ represents the propagation speed of waves in the medium. The term $v_{r}$ denotes the receiver's speed relative to the medium. If the receiver moves towards the source, $v_{r}$ is added to $c_{v}$; however, if the receiver moves away from the source, $v_{r}$ is subtracted. Similarly, $v_{s}$ represents the speed of the source relative to the medium. If the source moves away from the receiver, $v_{s}$ is added to $c_{v}$, and $v_{s}$ is subtracted if the source is moving towards the receiver.
In this work, we conceptualize humans (humans generating the response) or ChatGPT (ChatGPT generating the response) as sources and consider another human as the observer, as shown in Figure \ref{fig:do_ef}. The medium represents the space which variates or vibrates with the strength of the human factor in the input text embedding. Alternatively, the medium represents a space in which ChatGPT tries to imitate content similar to human-generated content. More detailed explanation is given in Section \ref{sub:des}.

\subsection{Energy-based model}
\label{subebm}
Statistical modeling and machine learning primarily aim to represent relationships between variables. By capturing these relationships, models can provide answers regarding the values of unknown variables based on the known variables. Energy-Based Models (EBMs) \cite{lecun2006tutorial}  assign "scalar values" to represent the compatibility between different variable configurations. These "scalar values" are called "energies" computed using an energy function defined to represent the compatibility between different variable configurations. Instead of focusing on classifying $\overrightarrow{x}$ values to corresponding $y$ values (as done in ML), the goal of EBMs is to determine whether a specific pair of ($\overrightarrow{x}, y$) is a good match or not. In other words, the objective is to find a compatible $y$ that fits well with a given $\overrightarrow{x}$. It can be framed as the task of discovering a $y$ for which a certain function $F(\overrightarrow{x}, y)$ has a low value, as the lower energy configurations describe the best relationship between these variables.
We establish an energy function $F:X \times Y \to \mathbf{R}$ that quantifies the degree of dependency between pairs $(\overrightarrow{x}, y)$. When constructing and training an energy-based model using a training set $S$, the process entails developing and refining four essential components:

\begin{itemize}[leftmargin=*]
    \item \textbf{The architecture: }The architecture of the EBM refers to the internal arrangement and design of the parameterized energy function $F(w,\overrightarrow{x},y)$ (with parameters $w$). It encompasses the specific structure and organization of the components within the energy function that are determined by the chosen model's parameters.
    \item \textbf{The inference algorithm: } The method for finding a value of $y$ that minimizes $F(w,\overrightarrow{x},y)$ for any given $\overrightarrow{x}$ is given by the following equation:
         \begin{equation}
            y^{'} = \argmin_{y} {F(w,x,y)} \nonumber
        \end{equation}
    \item \textbf{The loss function: } $L(w, F)$ quantifies the quality or performance of an energy function by evaluating its compatibility with the training set.
    \item \textbf{The learning algorithm: }The approach to discover a $w$ that minimizes the loss functional across the family of energy functions $F$, considering the given training set.
\end{itemize}

In this work, EBM is defined to model the dependencies between the input text embedding ($V(\overrightarrow{x})$) and the output label of the text. Label 1 for humans generating the text is assigned high energy, and label 0 for ChatGPT-generated responses is given low energy. The instantiation of all these components is detailed in Section \ref{sub:des}.

\subsection{Explainable AI}
\label{sub:xai}
Explainable AI (XAI), also known as interpretable AI or transparent AI, refers to the development and application of artificial intelligence (AI) systems that can effectively explain their reasoning and decision-making processes in a human-understandable manner. This has resulted in several novel model explanation techniques and toolkits \cite{ribeiro2016should, lundberg2017unified, shrikumar2017learning, arras2017relevant, jacovi2018understanding, ancona2018towards, GoogleMLInt, microsoftMLInt}. 
One class of XAI is the attributions method in which each feature contribution to the model function is computed. Mathematically, it is defined as: given an input data point $\overrightarrow{x} \in \mathbb{R}^n$ and a classification model $F$, an explanation method denoted as $\mathcal{H}$ is employed to provide an influence or attribution vector. This vector elucidates the model's decision-making process by revealing the contributions of each feature. The $i^{th}$ element of this vector, denoted as $\mathcal{H}_i(\overrightarrow{x})$, represents the impact of the $i^{th}$ feature on the predicted label $y$ for the given data point $\overrightarrow{x}$.
We utilize a specific attribution method, called Integrated Gradients (IG) \cite{sundararajan2017axiomatic}, which is derived by accumulating gradients computed at all points along a linear path from a chosen baseline $\overrightarrow{x}^{'}$ (often set as $\overrightarrow{x}^{'} = \overrightarrow{0}$) to the actual input $\overrightarrow{x}$ \cite{sundararajan2017axiomatic}. Essentially, integrated gradients involve integrating the gradients along a straight-line path from the baseline $\overrightarrow{x}^{'}$ to the input $\overrightarrow{x}$. The integrated gradient for a specific feature $i$ with an input $\overrightarrow{x}$ and baseline $\overrightarrow{x}^{'}$ is denoted as $\mathcal{H}_{IGRAD}(\overrightarrow{x})_i$, computed as shown in Equation \ref{eq:igrad}.

\begin{equation}
\mathcal{H}_{IGRAD}(\overrightarrow{x})_i = (x_{i} - x_{i}^{'}) * 
\int_{\alpha=0}^{1} \frac{\partial F(\overrightarrow{x} + \alpha (\overrightarrow{x} - \overrightarrow{x}^{'}))}{\partial x_i} \text{d}\alpha \label{eq:igrad}
\end{equation}

\noindent
In this work, we utilize IG to account for ChatGPT's rephrasing techniques that a human can request applied to evade detection from different ChatGPT detectors present in the literature. Specifically, we employ IG  to compute different perturbations of the input text embedding ($V(\overrightarrow{x})$) to extract crucial features. Then, we compute the energies of the perturbations by systematically removing these features, generating additional samples with corresponding labels, and re-optimizing the loss function. By integrating this in our proposed scheme, we efficiently improve our detection of ChatGPT-generated responses.

\begin{figure}[t]
\centering
\includegraphics[width=\linewidth]{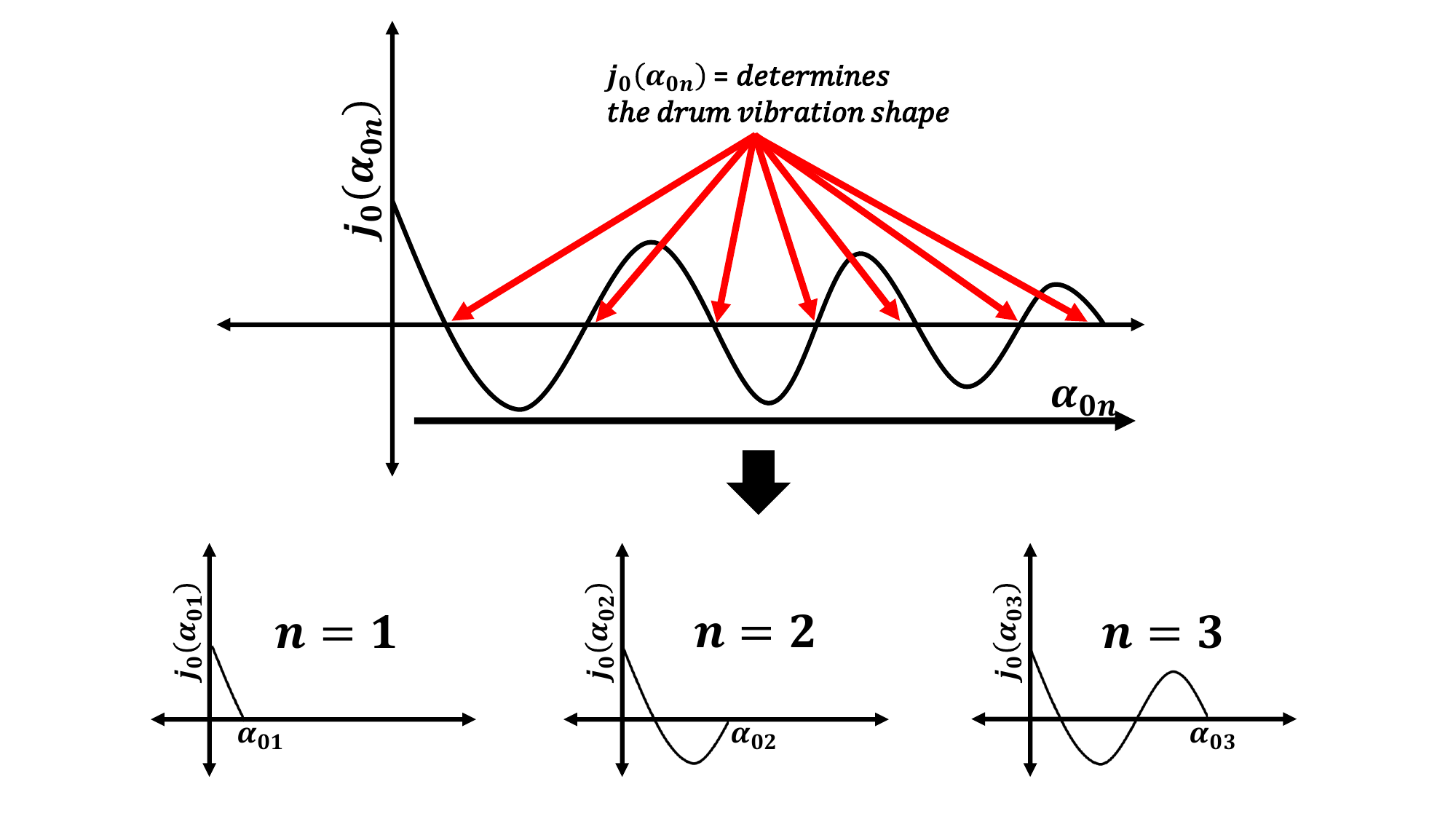}
\caption{The impact of different modes of the zeroth Bessel function ($J_{0} (\alpha_{0n})$) on the frequency with which the drumhead vibrates. For example, we have illustrated how the wave travels (shape of the wave) in a drum with a fixed $m=0$, and $n=1, \; n=2, \;n=3$. }
\label{fig:bessel}
\end{figure}

\subsection{Drumhead vibrations}
\label{sub:dv}
The behavior of an idealized drumhead can be understood by examining the vibrations of a two-dimensional elastic membrane under tension \cite{sapoval1991vibrations} \cite{jenkins2006membrane}. This membrane, which can be modeled as a circular surface with uniform thickness connected to a rigid frame, exhibits fascinating properties. The membrane can store vibrational energy at specific resonant frequencies through the resonance phenomenon. The surface of the membrane moves in distinct patterns characterized by standing waves, referred to as normal modes. The lowest frequency normal mode is called the fundamental mode, and the membrane possesses an infinite number of these normal modes. Specifically, there are two modes (that can have diverse values) of vibration in the membrane: angular mode ($m$) and circular mode ($n$).
The membrane can vibrate in countless ways, with each mode determined by the initial shape of the membrane and the transverse velocity at each point on the surface. The membrane's vibrations can be described mathematically by solving the two-dimensional wave equation, subject to Dirichlet boundary conditions that represent the constraints imposed by the frame \cite{vibrationsdrumhead}. This equation captures the intricate dynamics of the membrane and its interactions with the surrounding medium. The solution of this equation for asymmetric case ($m=0, n$), is given by:

\begin{equation}
\label{eq:drumhead}
    D_H(r,t) = \sum_{n=1}^{\infty} J_{0}(\lambda_{n} r) * (A\cos(c_{d} \lambda_{n} t) + B\sin(c_{d} \lambda_{n} t))
\end{equation}

$D_H(r, t)$ represents the height of the drum head at a specific radial coordinate $r$, measured relative to the "still" drum head shape at time $t$. Here, we have an open disk with a radius of $a$, where the value of $D_H$ is zero on the boundary. $c_{d}^{2}$ determines the wave speed in the membrane, in Equation~\ref{eq:drumhead} we use the root: $\sqrt{c_{d}^{2}} = c_{d}$. The coefficients $A$ and $B$ are determined by the initial conditions. $J_{0}(\lambda_{n} r)$ is the zeroth Bessel function with angular mode m=0, and circular mode n, and it represents the strength of the frequency computed at different modes for each $(0,n)$ value. The specific values of $\lambda_{n} r = \alpha_{n}$ are referred to as the zeros of the Bessel function $J_{0}$ \cite{bowman2012introduction} and are used to determine the shape of the drum that determines the frequency with which it vibrates, as shown in Figure \ref{fig:bessel}.
In this work, we utilize drumhead vibration to compute the source frequency (human or ChatGPT) ($E_{f_{0}}$) for the computation of observed frequency by an observer (human) in the Doppler effect.  We compute the zeros of the Bessel function $J_{0}(\alpha_{0n})$ (using the Python SciPy library \cite{bf}), which in turn is used to compute the source frequency. Frequency at mode $m=0,n$ can be computed as $\alpha_{0n} \times c_{d}/a$. Since, we are unaware of the values of $c_{d}$ and $a$, we compute $E_{f_{0}} = \frac{J_{0}(0, n)}{J_{0}(0, 1)}$ to normalize the computation of $E_{f_{0}}$ with fundamental frequency at $J_{0}(0, 1)$. Thus, we only need to measure $n$ to compute the source frequency (detailed in Section \ref{sec:design}). Also, we specifically concentrate on the circular mode of vibration. This choice is based on our assumption that the "Doppler waves" related to both human-generated and ChatGPT-generated content propagate in a linear fashion, or to put it differently, we assume that the source's waveform moves along the line of the observer's sight.
\section{Threat Model}
\label{sec:threat}
We investigate a situation where the detector or an observer (a human) functions as a black box and relies exclusively on the observed samples. However, the detector has no privileged knowledge about the underlying ChatGPT Large Language Model (LLM) responsible for generating these samples, including details such as its weights, structures, and gradients. This assumption accurately reflects the current situation, as OpenAI has not made the LLM utilized in the GPT-3.5 family publicly accessible, thereby limiting insider information. By considering this realistic scenario, we aim to explore the challenges and potential strategies of detectors against attacks in the absence of in-depth knowledge about the LLM. 
We make the following assumptions regarding the attacker: Firstly, we assume that the attacker is a human who actively aims to avoid detection. Secondly, the attacker has knowledge of the deployed detector.
However, the attacker has no information about our deployed detector's specific structural design, such as using EBM, doppler effect, and XAI. Additionally, we consider the attacker possibly rephrasing the given input string by utilizing ChatGPT multiple times to evade detection. The adversary may also employ other rephrasing techniques to alter the content and make it harder for the detector to identify. These assumptions acknowledge an attacker's potential strategies to bypass our detection system, prompting us to address the challenges associated with such deceptive practices.       
\section{System Design}
\label{sec:design}
In this section, we provide a comprehensive account of the primary constituents comprising our suggested ChatGPT detector. Initially, we present a broad overview of the concept, followed by a detailed description of its fundamental elements.

\begin{figure}[t]
\centering
\includegraphics[width=\linewidth]{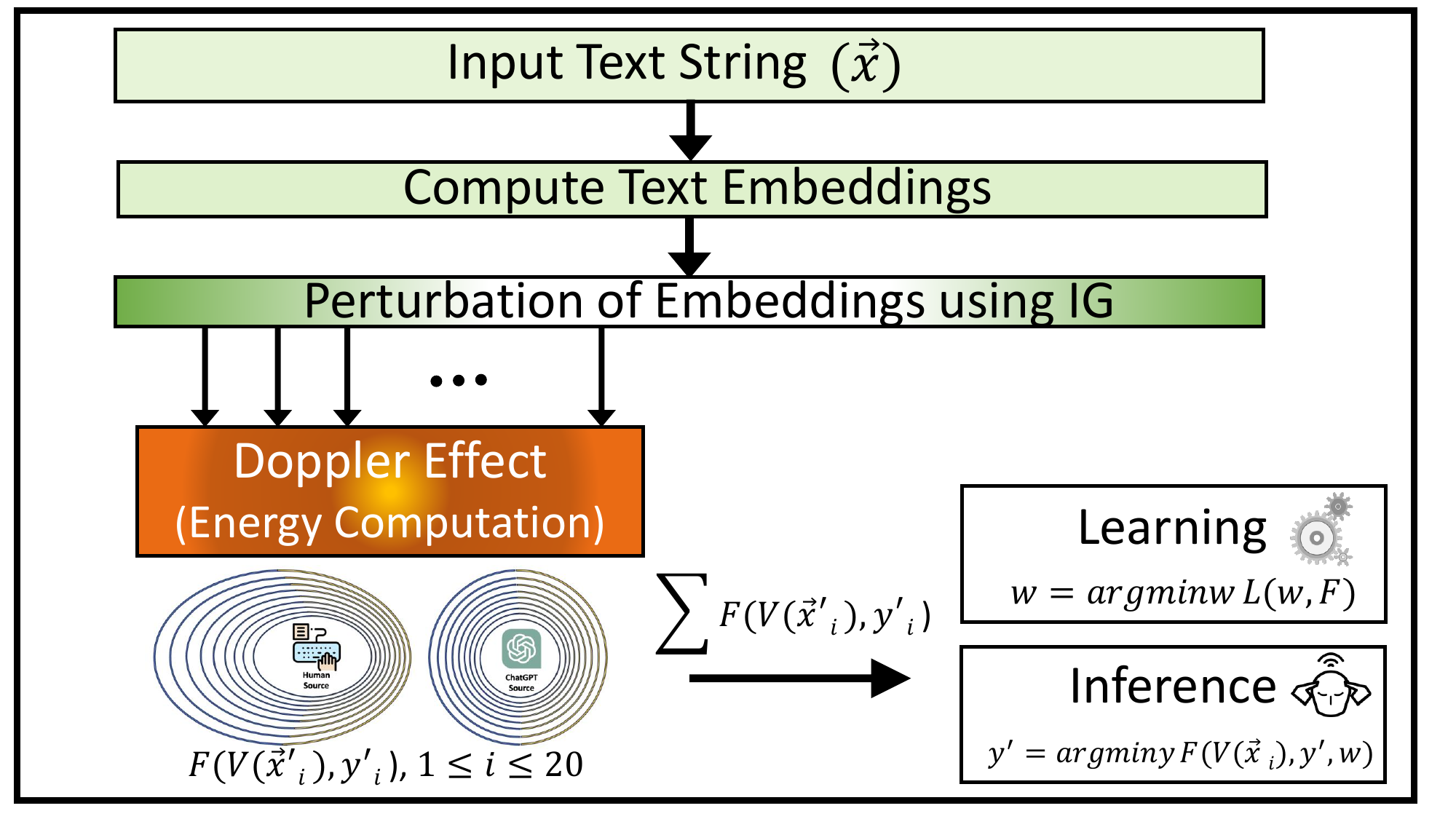}
\caption{High level idea of our proposed approach.}
\label{fig:highlevel}
\end{figure}

\subsection{High-level Idea}
\label{sub:hli}
Figure \ref{fig:highlevel} represents the high-level architecture of our proposed ChatGPT detector \ourname. 

As mentioned in the introduction, the main idea of our approach is to leverage the bias introduced by the human or ChatGPT in their responses to queries. Alternatively, one can perceive this as the fact that every individual interacts with others in a unique manner \cite{ingold2002culture, wykowska2015humans} and brings biases such as emotions, personal history with that specific person, and environmental factors, before engaging in communication.
Our approach consists of three components i) adapted Doppler effect, ii) energy-based model to capture the dynamics of human and ChatGPT behavior and iii) Integrated Gradient (IG) method to capture hybrid texts that contain both human- and ChatGPT-generate texts.     
We determine the different dynamics in human and machine behavior, by first building an Energy-Based Model (EBM) in which we quantify the energy using the Doppler effect. We assume that in the Doppler-effect context waves represent texts. During human communication, the recipient receives the phonetic energy contained in waveforms. Furthermore, the Doppler effect involves source waves radiating outward in concentric circles towards the observer, As a way to model source frequencies we incorporate drumhead vibrations. This decision is motivated by the fact that, similarly, diametric waves propagate in concentric circles towards a fixed outer boundary.

The optimization of the EBM training process is enhanced by incorporating the calculated wave frequencies into its cost function. This energy metric is employed to define the connection between the input text embedding and its corresponding output label (either 0 or 1). Hence, our approach completely differs from existing works that only propose a simple binary classifier ([34, 47]) trained on the input text.

To leverage the Doppler effect, we have made certain assumptions in our approach. Firstly, we consider the source of the text to be either a human or ChatGPT. If the source is a human, we assume that it is in motion with a velocity equal to the variance observed in responses generated by humans. Conversely, if the source is ChatGPT, we assume it is stationary. Additionally, we assume that the observer is another human and is moving with a constant velocity. Thus, we aim to determine how a human observer perceives a response generated by either a human or ChatGPT. To calculate the frequency of the source, we utilize the drumhead vibrations computed using Bessel functions $J_{0}(\alpha_{n})$, as described in Section \ref{sec:background}.
The reason for using drumhead vibrations is as follows: the Doppler effect involves source waves progressing toward the observer in concentric circles, and also, in the modelling of the vibrations (of waves) in drumhead, circular waves propagate in concentric circles toward the fixed outer boundary \cite{vibrationsdrumhead}.

Hence, the utilization of drumhead vibrations to calculate the source frequency is grounded in the understanding that waves, such as sound waves, propagate in concentric paths \cite{andrade1959doppler}.
There are two distinct modes of vibration associated with drumhead vibrations: circular ($n$) and angular ($m$). In this study, we specifically concentrate on the circular mode of vibration. This choice is based on our assumption that the "Doppler waves" related to both human-generated and ChatGPT-generated content propagate in a linear fashion, or to put it differently, we assume that the source's waveform moves along the line of the observer's sight.

Firstly, we compute the unique values in an embedding of the text ($V(\overrightarrow{x})$), and then employ the Bessel function ($J_{0}(\alpha_{n})$) to determine a specific source frequency ($E_{f_{\text{0}}}$), which is then further utilized to compute $E_{f}$. This is based on the idea that each human response vibrates at a distinct frequency. Furthermore, we employ explainable artificial intelligence (XAI) techniques, specifically Integrated Gradients (IG), to generate various perturbations of the input embedding. This is done to observe the behavior and outputs of different perturbed inputs from the perspective of an observer. We are motivated to understand how different perturbations of a particular input influence the perceived nature of the response. For instance, when a human rephrases a sentence using ChatGPT, the output should indicate that the response is generated by ChatGPT. However, if a human rephrases the same text multiple times, the output should indicate that the response is generated by a human. Thus, we aim to incorporate and integrate these effects into our methodology. To achieve this, we compute the energy associated with each perturbed text embedding and average it. Then, we integrate this averaged energy with Binary Cross Entropy (BCE) loss of the model function to have an effective total loss. Through optimization of this loss function, we are able to compute distinct energies for human-generated text and ChatGPT-generated text.

Furthermore, there are discernible distinctions between the energy of text embeddings and their associated labels across different classes (0 and 1) in the training process. Thus, it establishes a threshold between these classes that helps us determine whether the given input string is ChatGPT-generated or human-produced. Since, in DEMASQ, we have considered the count of perturbed embeddings (utilizing IG) that align with the original label, this count also indicates the shift from human-like responses to more ChatGPT-like responses the text evolves during rephrasing.

\subsection{\ourname Components Design and Implementation}
\label{sub:des}

Below, we present in detail the main components of our proposed ChatGPT detector \ourname. \\

\noindent
\textbf{Energy-based model.} We develop an energy-based model that quantifies a function known as "energy" to establish the relationship between $(V(\overrightarrow{x}), y)$, where $V(\overrightarrow{x})$ refers to the input text embedding, and $y$ refers to the output label $(0\;or\;1)$. We assign a label of $0$ to the ChatGPT-generated content and a label of $1$ to the human-generated response. In our design we employ the Sentence Transformers library along with the "msmarco-distilbert-base-tas-b" model to extract the embeddings of the input text. Subsequently, we construct a binary classification model with an energy function comprising the Binary Cross-Entropy loss (BCE) and a regularization energy term which computation involves employing the Doppler effect, as explained below.

To implement the model, we employ a neural network architecture consisting of six fully connected layers with Rectified Linear Unit (ReLU) activations. The sizes of the layers are set to 512, 256, 128, 64, 32, and 1, respectively. Additionally, we utilize the Adam optimizer with a learning rate of 0.0001 to train the model over a period of 12 epochs. Subsequently, we define the energy function to determine the relationship between $(V(\overrightarrow{x}), y)$ using the principles of the Doppler effect. We define the loss of EBM consisting the BCE loss and the energy loss $E_{f}$ (explained in the next section), mentioned below:\\

\begin{equation}
     loss = BCE + E_{f}(y)- min(E_{f}(\text{0}), E_{f}(\text{1}))
\end{equation}

\noindent where $E_{f}$ is the energy computed for true label, $E_{f}(0)$ is the energy computed for label 0 and $E_{f}(1)$ is the energy computed for label 1. The motivation behind incorporating the energy term is to incentivize the model to avoid making predictions that significantly diverge from the desired characteristics. Specifically, this means encouraging the model to assign labels 1 to instances with high-energy values and label 0 to instances with low-energy values.

\noindent
\textbf{Adapted Doppler Effect. }
As defined in Section \ref{sec:background}, Doppler effect is observed whenever the source of waves is moving with respect to an observer and is given by Equation \ref{eq:doppler}. 

Based on established principles, it is widely recognized that an object with a higher frequency possesses higher energy. Thus, we utilize the frequency of the observer, which is computed using the aforementioned formula, as the energy value for our energy-based model. To obtain the respective variables in Equation \ref{eq:doppler}, we perform the following computations:
$$v_{r} = 0.8$$
$$v_{s} = y \times abs(var(V(\overrightarrow{x})))$$
$$c_v = var(V(\overrightarrow{x})) \times J_{{0}}(0, 1)$$
The reasoning for determining the above values for the velocities of the source ($v_{s}$) and receiver ($v_{r}$) is as follows: considering that higher (lower) frequency corresponds to greater (lesser) energy and velocity, we establish that the frequency, energy, and velocity are directly proportional to each other. Frequency also indicates the rate of change of a specific value within a given timeframe. Therefore, to compute the source's velocity, we consider the variation of the input string embedding multiplied by its label. Thus, if the source corresponds to a human-generated response, represented by a label of 1, the source velocity ($v_{s}$) is computed as $1 \times abs(var(V(\overrightarrow{x})))$. However, for the ChatGPT-generated response, the computed source velocity ($v_{s}$) is zero ($0 \times abs(var(V(\overrightarrow{x})))$). Since we have perceived the receiver (observer) as human, we assume that it is moving with a constant velocity and assign a label of 0.8 (close to 1). Generally, in the Doppler effect, the speed of the medium, denoted as $c_v$, is constant (either the speed of sound or the speed of light.). Hence, in this paper, we integrate a constant of fundamental frequency $J_{0}(0,1)$ by multiplying it with the variance of $V(\overrightarrow{x})$. As detailed in Section \ref{sec:background}, the zeroes of $J_{0}(0,n)$ determine the shape of the drum that vibrates with a specific frequency; thus, $J_{0}(0,1)$ adds a constant of fundamental frequency (fixed size, as shown in Figure \ref{fig:bessel}) with which the medium vibrates and only variates with the strength of the human factor in $V(\overrightarrow{x})$ (computed using $var(V(\overrightarrow{x}))$). Alternatively, the medium represents a space in which ChatGPT tries to imitate content similar to human-generated content.

\begin{figure}[t]
\centering
\includegraphics[width=\linewidth]{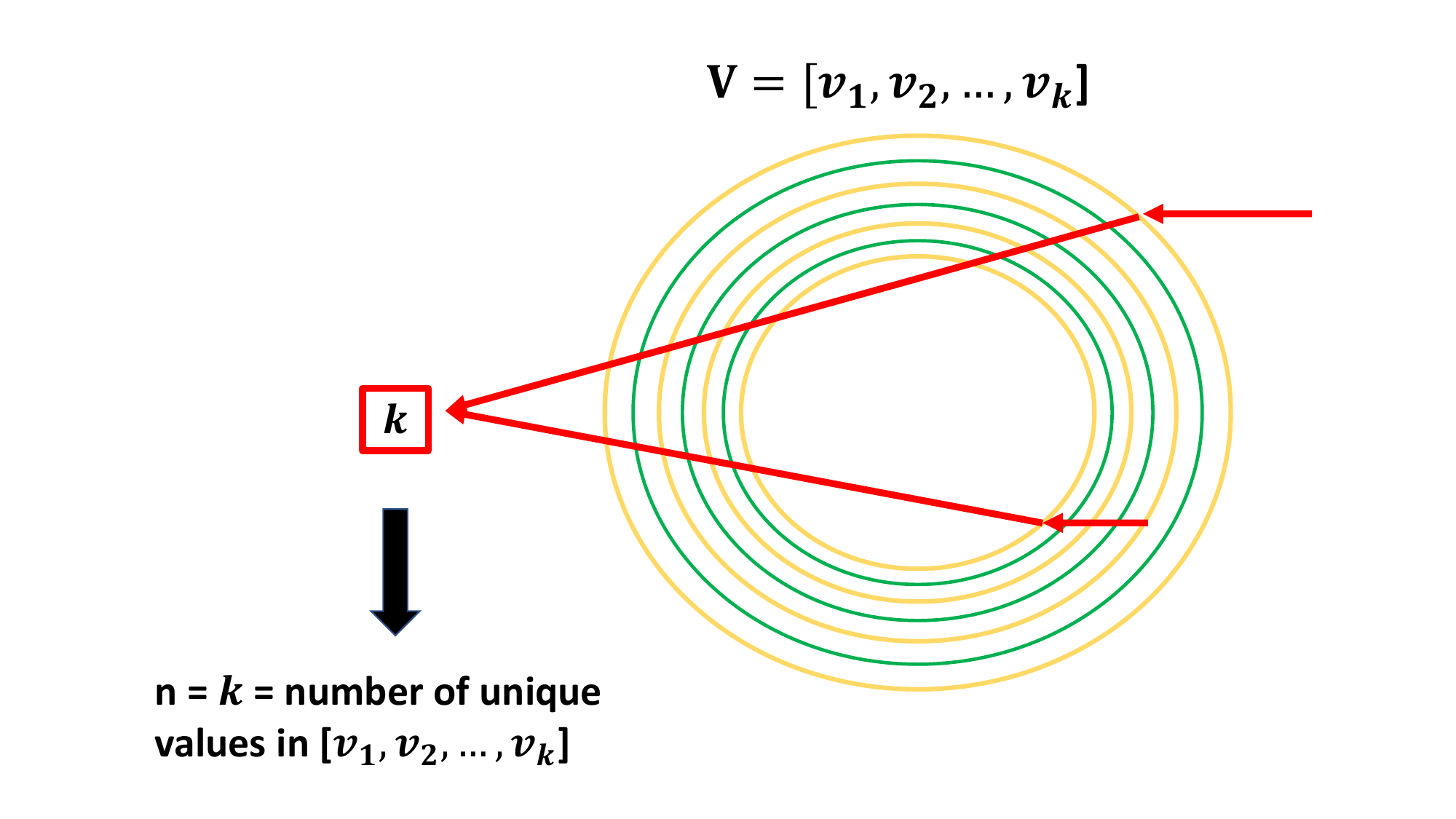}
\caption{The computation of the circular mode ($n$) to determine the frequency with which the drumhead vibrates. We compute the count of unique values ($k$) (unique concentric circles), as we have assumed each unique value in $V(\overrightarrow{x})$ as the radius of a concentric circle.}
\label{fig:drumhead}
\end{figure}

The following steps are executed to determine the source frequency $E_{f_{\text{0}}}$: as the source moves closer to or farther away from the observer, the frequencies are represented by concentric circles, as depicted in Figure \ref{fig:do_ef}. Our motivation is to determine the frequency associated with each embedding, $V(\overrightarrow{x})$. We achieve it by calculating the vibrating circular mode $(n)$ used in the Bessel function. We assume each value in $V(\overrightarrow{x})$ as the radius of a concentric circle, as shown in Figure~\ref{fig:drumhead}, but given that $V(\overrightarrow{x})$ can contain duplicate values, and we do not want to consider multiple circles with the same radius, our initial step involves computing the count of only distinct values in $V(\overrightarrow{x})$.

To simplify the computations and align the unique value’s of $V(\overrightarrow{x})$ with Euclidean plane’s origin ((0, 0)), the following steps are undertaken: if the highest or lowest value in $V(\overrightarrow{x})$ is less than 0, that value is added to each value in $V(\overrightarrow{x})$. This adjustment ensures that the center of the circles is located at $(0,0)$. Alternatively, if the lowest positive value is greater than or equal to 0, it is appended to each value in $V(\overrightarrow{x})$ to achieve the same result. Consequently, the centers of the circles (representing frequencies) are shifted to the origin. However, this shift does not affect our methodology because we only require the unique values of the adjusted embedding to compute the source frequency using the Bessel function, as explained below in the description of Algorithm 1.

Algorithm \ref{algo:sfc} and Algorithm \ref{algo:drumf} encompass all these steps, enabling the computation and retrieval of the source frequency, whether it pertains to a human or ChatGPT. Algorithm \ref{algo:sfc} computes the source frequency of $V(\overrightarrow{x})$, which is assumed to have length $s$. On line 3, it computes the minimum and maximum values of the sequence, denoted as $min_{V(\overrightarrow{x})}$ and $max_{V(\overrightarrow{x})}$, respectively. Next on line 4, 5, and 6, it checks if the minimum value, $min_{V(\overrightarrow{x})}$, is less than 0. If it is, it implies that the sequence contains negative values. In this case, the algorithm shifts the entire sequence by adding $min_{V(\overrightarrow{x})}$ to each element, ensuring that the minimum value becomes 0. Else, on line 8 and line 9, if the minimum value is greater than or equal to 0, the algorithm assumes there are no negative values in the sequence. In this case, it shifts the sequence by subtracting $max_{V(\overrightarrow{x})}$ from each element, making the maximum value become 0. Then, on line 10, it calculates the count of the unique values in the updated sequence, denoted as $u_{V(\overrightarrow{x})}$. These unique values represent the distinct concentric circles present in the sequence, as shown in Figure \ref{fig:drumhead}. On line 11, it calls a function, $\texttt{computezeroes}$, passing 0 and the length of $u_{V(\overrightarrow{x})}$. The reason we pass 0, which is the angular node $(m)$ in Bessel functions, is because we assume that the source (human-generated or ChatGPT-generated content) is moving linearly towards the observer, as explained in Section \ref{sub:dv}.
Finally, the computed source frequency ($E_{f_{0}}$) is returned as the output of the algorithm. \\

\begin{algorithm}[tb]
\caption{Source Frequency Computation ($\texttt{SourceFrequency}$), $E_{f_{0}}$}
\label{algo:sfc}
\begin{algorithmic}[1]
    \State \textbf{Input: $V(\overrightarrow{x})$ having length $s$ }
    \State \textbf{Output: Source Frequency }
    \State Compute $min_{V(\overrightarrow{x})} = \texttt{min}(V(\overrightarrow{x}))$ and $max_{V(\overrightarrow{x})} = \texttt{max}(V(\overrightarrow{x}))$
        \If{$min_{V(\overrightarrow{x})} < 0$}
            \For{$i=1$ to $n$}
                \State $V(\overrightarrow{x})[i] = V(\overrightarrow{x})[i] + min_{V(\overrightarrow{x})}$
            \EndFor
        \Else
            \For{$i=1$ to $n$}
                \State $V(\overrightarrow{x})[i] = V(\overrightarrow{x})[i] - max_{V(\overrightarrow{x})}$
            \EndFor
        \EndIf
        \State $u_{V(\overrightarrow{x})}$ = unique values in $V(\overrightarrow{x})$
        \State source frequency = $\texttt{computezeroes}(0,\texttt{len}(u_{V(\overrightarrow{x})}))$
        
    \State \Return source frequency, $E_{f_{0}}$.
\end{algorithmic}
\end{algorithm}

\begin{algorithm}[tb]
\caption{Compute Drumhead Frequency ($\texttt{computezeroes}$)}
\label{algo:drumf}
\begin{algorithmic}[1]
    \State \textbf{Input: Number of diametric nodes ($0$), Number of circular nodes ($c_n$)}
    \State \textbf{Output: Drumhead Frequency}
    \State Set $d_{\texttt{max}} = m = 0$ \Comment{Number of diametric nodes to calculate}
    \State Set $c_{\texttt{max}} = n = c_n$  \Comment{Number of circular nodes to calculate}
    \State Compute drumhead fundamental frequency: $J_{0}(0, 1)$
    \State Drumhead frequency $\gets \frac{J_{0}(m, n)}{J_{0}(0, 1)}$
    \State \Return Drumhead frequency
\end{algorithmic}
\end{algorithm}

\noindent
\textbf{Drumhead vibrations.} In Algorithm \ref{algo:drumf}, we compute the source frequency using the Drumhead vibrations that are computed utilizing the Bessel functions, as given in Section \ref{sec:background}. Specifically, we use ($J_{0}(0,n) / J_{0}(0,1)$) to compute the relative frequency for different zeroes of $J_{0}(0,n)$ with respect to fundamental $J_{0}(0,1)$ mode frequency. First, we set the number of diametric nodes, i.e., $m$ equal to zero and the number of circular nodes, i.e., $n$ equal to the number of unique values in $V(\overrightarrow{x})$ and pass it to the Bessel function to compute the frequency at that particular mode of the Bessel function. Algorithm \ref{algo:drumf} proceeds as follows: the inputs to the algorithm are the number of diametric nodes and the number of circular nodes. On line 3 and line 4, we set the input parameters. The number of diametric nodes, denoted as $d_{\texttt{max}}$, is set to zero, and the number of circular nodes, denoted as $c_{\texttt{max}}$ ($n$), is set to the value of the input parameter $c_n$, which is determined as explained before. On line 6, we compute the drumhead fundamental frequency. This frequency is obtained by evaluating the Bessel function $J_{0}(0, 1)$. Finally on line 7, the drumhead frequency is calculated, by dividing the value of $J_{0}(m, n)$ by the value of $J_{0}(0, 1)$. The algorithm returns the computed drumhead frequency as the output.
In summary, the algorithm takes the number of diametric and circular nodes as input, computes the drumhead fundamental frequency, and then calculates the drumhead frequency based on the obtained fundamental frequency. The resulting drumhead frequency is returned as the output of the algorithm. \\ 

\noindent
\textbf{Computing Perturbations of the embedding text.} Our second objective, as stated in Section \ref{sec:introduction} and Section \ref{sec:background}, is to accurately determine whether the given text is produced by ChatGPT, even when it undergoes multiple human manipulations to avoid detection. To achieve this, we employ the XAI IG method to perturb the input text embedding, as shown in Section \ref{sec:background}. We set the baseline $V(\overrightarrow{x}^{'}) = \overrightarrow{0}$ because \cite{sundararajan2017axiomatic} shows inaccurate gradient attribution maps for other baselines. Initially, we identify the top 20 most relevant features, which are the features that contribute the most to the output label of the input text embedding. In other words, we identify the features that make the given content ChatGPT or human. Subsequently, we systematically set each relevant feature to zero, one at a time, to obtain the perturbed input. This approach allows us to generate a perturbed input text embedding that leverages the most significant features from the original text embedding. Finally, we calculate the energies (or frequencies) for each perturbed sample using Algorithm \ref{algo:sfc} and \ref{algo:drumf}, average them, and utilize the aggregated energy to optimize the loss function.

\subsection{Detecting ChatGPT-generated content}
\label{subsec:algo}
Algorithm \ref{algo:mainalgo} aims to train and test an Energy-Based Model (EBM) to distinguish between ChatGPT-generated content and human-generated content. Below, we explain in detail it's inner working. It takes as input a sequence, denoted as $V(\overrightarrow{x})$, with length $s$, and its corresponding true label, $y$, which indicates whether the content is ChatGPT-generated or human-generated. It also receives the number of epochs, $T$, for training the model. The output of the algorithm is the model's prediction, which can be either 0 or 1. A prediction of 0 represents ChatGPT-generated content, while a prediction of 1 indicates human-generated content. On line 3, it enters a loop that iterates from $epoch=1$ to $T$ for training the model. Inside this training loop, we compute the observer's frequency, $E_{f}$ (line 8), by applying the Doppler effect. This computation involves considering the speed of the medium, $c$ (line 4), which is calculated based on the variance of the input sequence, $V(\overrightarrow{x})$. Additionally, the speed of the source, $v_{s}$ (line 5), is determined by multiplying the true label, $y$, with the absolute value of the variance of $V(\overrightarrow{x})$. The speed of the receiver, $v_{r}$ (line 6), is set to a fixed value of 0.8. The source frequency, $E_{f_{\text{0}}}$ (line 7), is obtained by using the function $\texttt{SourceFrequency}(V(\overrightarrow{x}))$ on the input embedding $V(\overrightarrow{x})$. The observer's frequency, $E_{f}$ (line 8), is computed as $\left(\frac{c + v_{r}}{c - v_{s}}\right)E_{f_{\text{0}}}$. This calculation considers the effects of the speeds of the medium, source, and receiver, as well as the source frequency. For training the EBM, on line 9, we set the loss function. The loss is defined as a combination of two terms: binary cross-entropy (BCE) loss and $ E_{f}(y)- min(E_{f}(0), E_{f}(1))$ (Section \ref{sub:des} \textbf{Energy-based model}). This choice of loss function aims to optimize the model's performance in differentiating between ChatGPT-generated and human-generated content. The Adam optimizer \cite{kingma2017adam} is selected for training the model. It is a popular optimization algorithm widely used in machine learning. The loop continues for the specified number of epochs, during which the model is trained by adjusting its parameters to minimize the defined loss function.

The algorithm assigns negative energies for ChatGPT-generated content and positive energies for human-generated content. This step involves calculating and assigning energies based on the generated content. Finally on line 12, the algorithm returns the energy of the test query, which represents the model's prediction of whether the given content is generated by ChatGPT or a human.

In summary, the algorithm iteratively trains an EBM model using the Doppler effect-based observer's frequency and a combination of BCE loss and the observer's frequency as the training objective. The trained model is then utilized to predict whether a given content is ChatGPT-generated or human-generated based on the assigned energies.

\begin{algorithm}[tb]
\caption{\ourname{} main working.}
\label{algo:mainalgo}
\begin{minipage}{1.12\columnwidth}
\begin{algorithmic}[1]
    \State \textbf{Input:} $\overrightarrow{x}$ with length $s$, true label $y$, training epochs $T$
    \State \textbf{Output:} Model output: 0 or 1. "0" for ChatGPT-generated content and "1" for human-generated content
    \For{$epoch=1$ to $T$}
        \State Speed of the medium, \textit{c} = var($\overrightarrow{x}$)
        \State Speed of the source, $v_{r}$ = $y \times \texttt{abs(var}(\overrightarrow{x}))$
        \State Speed of the receiver, $v_{s}$ = 0.8
        \State Source frequency, $E_{f_{\text{0}}}$ = \texttt{SourceFrequency}($\overrightarrow{x}$)
        \State Observer frequency, $E_{f}$ = $\left(\frac{c + v_{r}}{c - v_{s}}\right)E_{f_{\text{0}}}$ \Comment{Computing Observer's frequency (\textit{$E_{f}$}) using Doppler effect.}
        \State Set EBM loss function: loss = BCE +  $E_{f}(y)- min(E_{f_{\text{0}}},\; E_{f_{\text{1}}})$ \Comment{Training EBM Model.}
        \State Set Optimizer = Adam(lr: 0.0001)
    \EndFor
    \State Negative energies for ChatGPT-generated content and positive for human-generated content  \Comment{Testing EBM Model.}
    \State \textbf{Return the energy of the test query.}
\end{algorithmic}
\end{minipage}
\end{algorithm}

\section{Experimental Setup}
\label{sec:eval_setup}

\subsection{Setup}
\label{setup}
Our experiments are conducted using the PyTorch~\cite{pytorch} framework on a server equipped with 4 NVIDIA RTX 8000 (each with 48GB memory), an AMD EPYC 7742, and 1024 GB of main memory. To evaluate the efficacy of our detector, we utilize a benchmark dataset (Section \ref{sub:dataset}) containing prompts from both ChatGPT and humans, covering a wide range of domains such as Medical, Wikipedia, Open
Q\&A, Reddit, Finance, arXiv, and Political (cf. Section \ref{sec:introduction}). In the subsequent section, we detail the configurations of the different datasets and the accuracy and precision metrics utilized to evaluate the performance of our approach, \ourname.

\subsection{Benchmark Dataset}
\label{sub:dataset}

Our benchmark dataset is created using a variety of datasets spanning multiple domains, as mentioned above. To generate the samples from ChatGPT, we made use of the queries gathered by Guo et al.~\cite{guo2023close} that also presented the corresponding human answer, then through the use of OpenAI's API\footnote{\url{https://openai.com/blog/introducing-chatgpt-and-whisper-apis}} we queried the latest version of GPT-3.5 at this time (gpt-3.5-turbo) requesting 3 different answers for each input question. Due to the limitation of the available data and the reported unstable behavior of ChatGPT~\cite{openai2023gpt4}, we assumed that the API would properly perform only with English prompts, therefore only querying for answer in this language.
The dataset comprises a total of 134,178 unique samples, with 59,214 responses generated by humans and 74,964 responses generated by the ChatGPT model. These samples cover 25,290 distinct questions from various domains, including medicine, open-domain, finance, politics, and rephrased ArXiv's abstract to cover the hybrid text case. Moreover, we expanded the dataset by incorporating responses from popular social networking platforms (such as Reddit and Wikipedia Q\&A), offering a diverse range of user-generated perspectives.
\noindent To ensure that the answers generated were diverse and not simply reworded versions of the same response, we employed the \mbox{\textit{msmarco-distilbert-base-tas-b}}\footnote{\url{https://huggingface.co/sentence-transformers/msmarco-distilbert-base-tas-b}} sentence transformer to asses that those responses had a low grade of sentence similarity. Lastly, we observed that the human samples provided by Guo et al.~\cite{guo2023close} contained artifacts and symbols, probably due to flaws in the scraping algorithm used to obtain the text. Similarly, in the response of ChaGPT there were also artifacts and symbols, probably intended to implement formatting or symbols to its response (such as table tags or bullet points). Therefore, we employed the sentence transformer \mbox{\textit{all-MiniLM-L6-v2}}\footnote{\url{https://huggingface.co/sentence-transformers/all-MiniLM-L6-v2}} to check and remove the artifacts from those samples. The benchmark dataset we constructed serves as a standardized reference for evaluating the effectiveness of different techniques in detecting ChatGPT-generated content.

\subsection{Evaluation Metrics} 
In order to assess and compare the effectiveness of our approach, we employed the following metrics:
\begin{itemize}
\item \textit{True Positive Rate (TPR): } This metric represents the tool's sensitivity in detecting text that ChatGPT generates. True Positive ($TP$) is the total number of correctly identified samples, while we consider False Negative ($FN$) the number of samples not classified as generated text or incorrectly identified as human text. Therefore, $TPR = \frac{TP}{TP+FN}$.
\item \textit{True Negative Rate (TNR):} This metric indicates the tool's specificity in detecting human-generated texts. True Negatives ($TN$) is the total number of correctly identified samples, while False Positives ($FP$) is the number of samples incorrectly classified as being produced by ChatGPT. Therefore, $TNR = \frac{TN}{TN+FP}$.
\end{itemize}
\section{Experimental Evaluations}
\label{sec:eval}

\begin{table*}[t]
\centering
\resizebox{0.8\textwidth}{!}{
\renewcommand{\arraystretch}{1.1}
\begin{minipage}{\textwidth}
    \caption{Summary of analyzed papers. TPR represents the detection capability of ChatGPT-generated text, while TNR reflects the detection capability of Human-generated text.}
    \label{tab:summary}
    \centering
        \large
\center
\begin{tabular}{|l|c|c|c|c|c|}
\hline
\multicolumn{1}{|c|}{\multirow{2}{*}{Approach}} & \multicolumn{1}{c|}{\multirow{2}{*}{\begin{tabular}[c]{@{}c@{}}Published\\ in\end{tabular}}} & \multirow{2}{*}{\begin{tabular}[c]{@{}c@{}}Publicly\\ Available\end{tabular}} & \multirow{2}{*}{Free/Paid} & \multirow{2}{*}{TPR (\%)} & \multirow{2}{*}{TNR (\%)} \\
\multicolumn{1}{|c|}{} & \multicolumn{1}{c|}{}  &                &                   &                          &    \\ \hline
Bleumink et al.~\cite{bleumink2023keeping}  & \multicolumn{1}{c|}{2023}                               & \checkmark                                                                           & Paid                   & 13.4                                                                                     & 95.4                                                                                        \\ \hline
ZeroGPT~\cite{AITextDetector}          & \multicolumn{1}{c|}{2023}                               & \checkmark                                                                           & Paid                   & 45.7                                                                                     & 92.2                                                                                        \\ \hline
OpenAI  Classifier~\cite{AITextClassifier}  & \multicolumn{1}{c|}{2023}                            & \checkmark                                                                           & Free                  & 31.9                                                                                     & 91.8                                                                                        \\ \hline
GPTZero~\cite{GPTZero}           & \multicolumn{1}{c|}{2023} & \checkmark                                                                           & Paid                   & 27.3                                                                                  
& 93.5                                                                                    
\\ \hline
Hugging Face~\cite{Huggingface}      & \multicolumn{1}{c|}{2023}                              & \checkmark                                                                           & Free                   & 10.7                                                                                      & 62.9                                                                                        \\ \hline
Guo et al.~\cite{guo2023close}  & \multicolumn{1}{c|}{2023} & \checkmark                                                                           & Free                   & 47.3                                                                                     
& 98.0
\\ \hline
Perplexity (PPL)~\cite{perplexityAnalysis} & \multicolumn{1}{c|}{2023}     & \checkmark                                                                           & Free                   & 44.4                                                                                    
& 98.3
\\ \hline
Writefull GPT~\cite{writefull}  & \multicolumn{1}{c|}{2023}          & \checkmark                                                                           & Paid                   & 21.6                                                                                      & 99.3                                                                                        \\ \hline
Copyleaks~\cite{copyleaks}       & \multicolumn{1}{c|}{2023}              & \checkmark                                                                           & Paid                   & 22.9                                                                                      & 92.1                                                                                        \\ \hline
Cotton et al.~\cite{cotton2023chatting}     & \multicolumn{1}{c|}{2023}        & $\times$                                                                         &           -                 & -                                                                                         & -                                                                                            \\ \hline
Khalil et al.~\cite{khalil2023will}     & \multicolumn{1}{c|}{2023}                 & $\times$                                                                         &            -                & -                                                                                         & -                                                                                            \\ \hline
Mitrovic et al.~\cite{mitrovic2023chatgpt}    & \multicolumn{1}{c|}{2023}                      & $\times$                                                                           &               -             & -                                                                                         & -                                                                                            \\ \hline

Content at Scale~\cite{aicontentdetector}  & \multicolumn{1}{c|}{2022}                       & \checkmark                                                                           & Paid                   
& 38.4  & 79.8
\\ \hline

Orignality.ai~\cite{originality}  & \multicolumn{1}{c|}{2022}                      & $\times$                                                                          & Paid                   & 7.6                                                                                     & 95.0                                                                                        \\ \hline

Writer AI Detector~\cite{WriterAIContentDetector}   & \multicolumn{1}{c|}{2022}         & \checkmark                                                                           & Paid                   
& 6.9  & 94.5
\\ \hline

Draft and Goal~\cite{DAG}        & \multicolumn{1}{c|}{2022}                 & \checkmark                                                                           & Free                  & 23.7                                                                                      & 91.1                                                                                        \\ \hline

Gao et al.~\cite{gao2022comparing}    & \multicolumn{1}{c|}{2022}                    & $\times$                                                                           &           -                 & -                                                                                         & -                                                                                            \\ \hline

Liu et al.~\cite{liu2023check} (Detector 1 - Task1)   & \multicolumn{1}{c|}{2023}                    & \checkmark                                                                           &  Free                 & 2.1                                                                                         & 98.8                                                                                            \\ \hline

Liu et al.~\cite{liu2023check} (Detector 2 - Task2)   & \multicolumn{1}{c|}{2023}                    & \checkmark                                                                           &  Free                 & 13.3                                                                                         & 99.4                                                                                            \\ \hline
Liu et al.~\cite{liu2023check} (Detector 3 - Task3)   & \multicolumn{1}{c|}{2023}                    & \checkmark                                                                           &  Free                 & 6.6                                                                                         & 98.7                                                                                           \\ \hline
\end{tabular}
\end{minipage}
}
\end{table*}

Next, we empirically illustrate the effectiveness of \ourname against the curated dataset and compare its efficacy against various baseline detection mechanisms. Further, we show how energy computation varies for the human and the ChatGPT-generated responses. Finally, we demonstrate the robustness of \ourname against the rephrasing techniques and transferability of \ourname to other domains, i.e., different from the domains that our model is trained on. 

\subsection{Baseline Approaches}
\label{sub:baseline}
In our evaluation, we assessed various tools and algorithms that have been proposed so far to determine their effectiveness in detecting ChatGPT-generated content, as summarized in Table \ref{tab:summary}. This table provides an overview of these tools' detection capabilities, specifically in terms of true positive rate (TPR) and true negative rate (TNR), when applied to ChatGPT-generated text. We considered several attributes of the tools mentioned in the literature for detecting ChatGPT-generated content, including the year of release, availability (public or private), and cost (free or paid). Our analysis revealed that none of the evaluated approaches consistently achieved high detection rates for ChatGPT-generated text. The most effective online tool we examined demonstrated a success rate of less than 50\%, as depicted in Table \ref{tab:summary}.
Notably, the maximum accuracy achieved by the detector was around 47\%. In addition, recently the authors of \cite{liu2023check} propose a ChatGPT-detector, CheckGPT, for academic abstracts of research papers. It achieves an accuracy of 98\% to 99\%. Nevertheless, their detector's performance significantly declined, as we evaluated it against our benchmark dataset. Thus, it demonstrates the limited effectiveness of their approach, with accuracy ranging from 2.1\% to 13.3\%, as further detailed in Section \ref{sub:rephrase}. \\
We believe that our dataset serves as a reliable benchmark to evaluate their approach since it contains 134,178 diverse and balanced samples, effectively measures the similarity between human and ChatGPT responses, and leverages responses from various domains and social networks. As a result, we regard all the examined detectors as baselines for our study. In contrast, our proposed approach, \ourname, substantially surpasses existing detection methods, achieving an accuracy ranging from 74.5\%.

\subsection{Discipline-Specific detection}
\label{sub:discipline}
This section illustrates our detection approach against different sub-discipline datasets that constitute our benchmark dataset. Sub-disciplines dataset includes samples from different domains such as Medical, Wikipedia, Open Q\&A, Reddit, Finance, arXiv, and Political. For each of these sub-datasets, we individually train our EBM (described in Section \ref{sec:design}) and analyze the energy spread for the human-generated responses and ChatGPT-generated responses. Table \ref{tab:subdatasetsenergy} details our training and inference results. Since we are training different EBM models for different datasets, we are designing different detectors for different fields. Thus, our work provides a generic detector that can be deployed in any field to detect whether the text is ChatGPT or human-generated.

Table \ref{tab:subdatasetsenergy} provides an overview of the performance of different sub-datasets in the conducted experiments. For a specific dataset, we mention the number of samples in each dataset, the True Positive Rate (TPR), and the True Negative Rate (TNR). Next, we examine the findings of these specific datasets to illustrate the insights obtained. The Medical dataset comprises 4,992 records, achieving a TPR of 96.8\% and a TNR of 96.4\%. It indicates that the model trained on the Medical dataset exhibits an efficacy of 96.8\% in correctly identifying ChatGPT-generated responses and 96.4\%  in correctly identifying human-generated responses.
Similarly, the Financial dataset, consisting of 15,732 records, demonstrates a TPR of 93.6\% and a TNR of 93.7\%. With 3,368 records, the Wiki dataset achieves a TPR of 84.2\% and a TNR of 83.2\%. The Open Q\&A dataset, encompassing 4,748 records, shows a TPR of 74.5\% and a TNR of 73.0\%. Among the examined datasets, the Reddit dataset stands out with 99,054 records and perfect scores of 100\% for both TPR and TNR. This analysis implies that the model trained on the Reddit dataset accurately identifies ChatGPT-generated and human-generated content.
Furthermore, the Political dataset, consisting of 3,620 records, demonstrates a TPR of 92\% and a TNR of 91.5\%. The arXiv dataset, comprising 2,664 records, shows a TPR of 90\% and a TNR of 89.5\%. Finally, the Combined dataset, consisting of 134,178 records, achieves a TPR of 97\% and a TNR of 96.5\%. Overall, each dataset exhibits varying levels of accuracy, as reflected in their respective TPR and TNR values. These results shed light on the models' performance on different datasets, allowing for a comparative analysis of their ability to accurately identify ChatGPT-generated and human-generated responses.

Further, we trained our model using the Task1, Task2, and Task3 academic abstract datasets provided by Liu et al. \cite{liu2023check}, as detailed in Table \ref{tab:checkgpt_subdatasetsenergy}. In their research \cite{liu2023check}, the authors developed CheckGPT, a detector for LLM-generated content, specifically designed to determine if academic abstracts in CS, physics, and humanities and social sciences (HSS) were authored by ChatGPT or not. Liu et al. \cite{liu2023check} broke down the analysis of CheckGPT into three tasks: 1) Full abstracts written by GPT model, 2) Abstracts partially completed by GPT model, and 3) Abstracts polished by GPT model. For Task1, the researchers presented ChatGPT with a title and asked it to generate a corresponding abstract. For Task2, they supplied ChatGPT with the first part of an abstract and requested it to complete the rest, with the word count 'w' equalling the number of words in the second half of in the original abstract. For Task3, the researchers gave ChatGPT a complete abstract to enhance. We use CS-TASK datasets that the authors made available and evaluated \ourname against all these tasks. Our \ourname EBM was retrained on Task1, Task2, and Task3 datasets, demonstrating accuracies of 96.4\% for Task1, 88.7\% for Task2, and 82.5\% for Task3.

\begin{figure*}[t]
    \centering
    \begin{tabular}{cccc}
        \begin{subfigure}[b]{0.24\textwidth}
            \centering
            \includegraphics[width=0.75\textwidth]{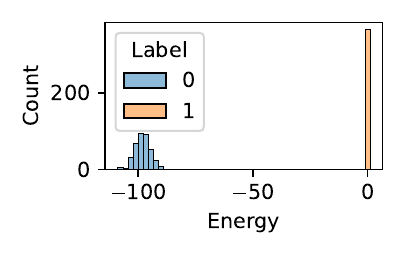}
            \caption{Medical}
            \label{fig:medical_hist}
        \end{subfigure}
        &
        \begin{subfigure}[b]{0.24\textwidth}
            \centering
            \includegraphics[width=0.7\textwidth]{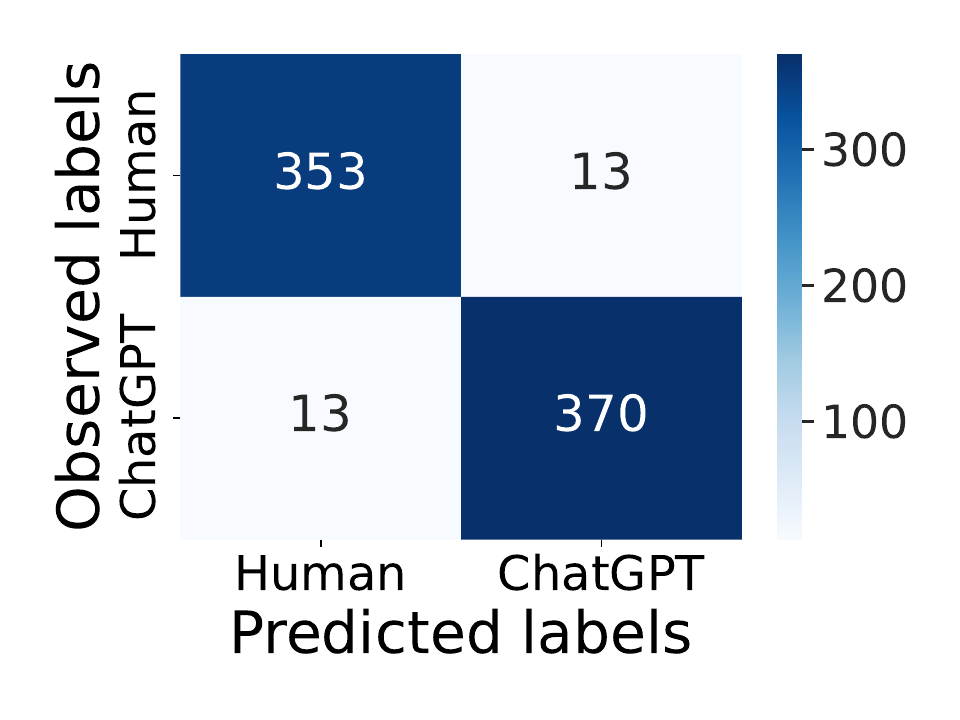}
            \caption{Medical}
            \label{fig:medical_scatter}
        \end{subfigure}
        &
        \begin{subfigure}[b]{0.24\textwidth}
            \centering
            \includegraphics[width=0.75\textwidth]{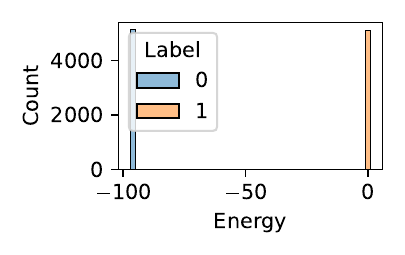}
            \caption{Reddit}
            \label{fig:reddit_hist}
        \end{subfigure}
        &
        \begin{subfigure}[b]{0.24\textwidth}
            \centering
            \includegraphics[width=0.7\textwidth]{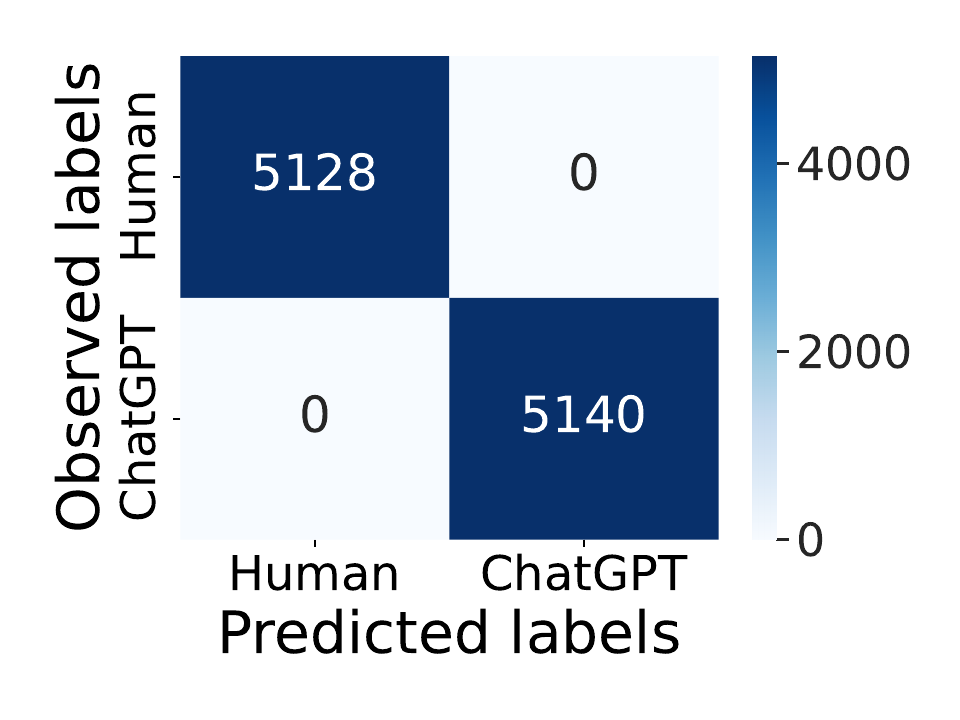}
            \caption{Reddit}
            \label{fig:reddit_scatter}
        \end{subfigure}
        \\
        \begin{subfigure}[b]{0.24\textwidth}
            \centering
            \includegraphics[width=0.75\textwidth]{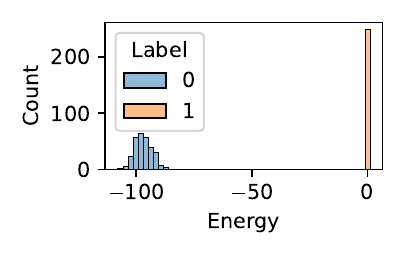}
            \caption{Political}
            \label{fig:political_hist}
        \end{subfigure}
        &
        \begin{subfigure}[b]{0.24\textwidth}
            \centering
            \includegraphics[width=0.7\textwidth]{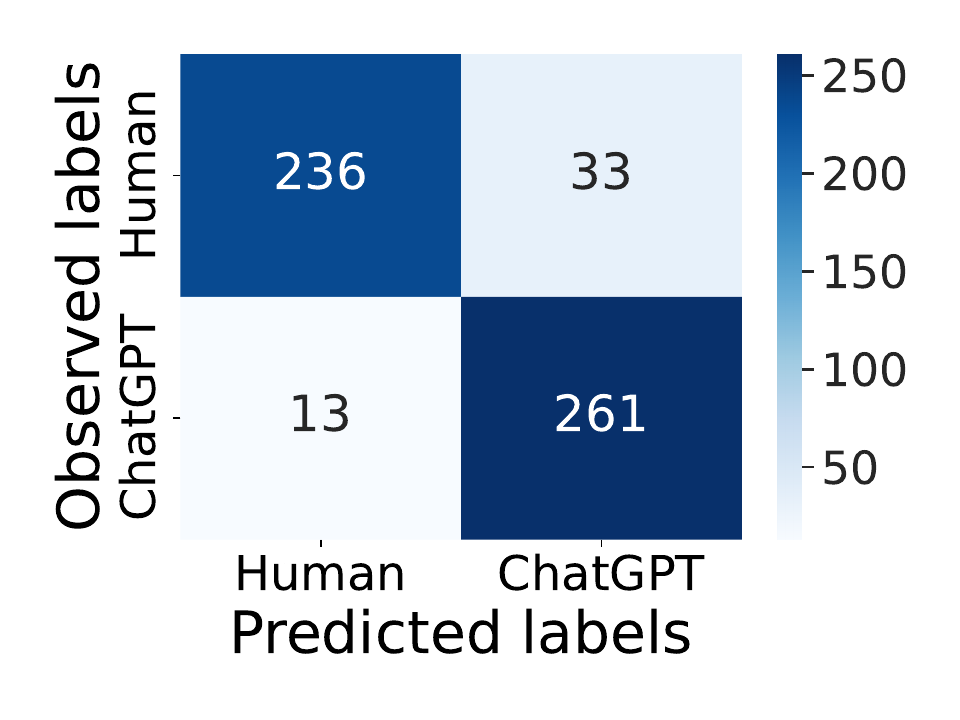}
            \caption{Political}
            \label{fig:political_scatter}
        \end{subfigure}
        &
        \begin{subfigure}[b]{0.24\textwidth}
            \centering
            \includegraphics[width=0.75\textwidth]{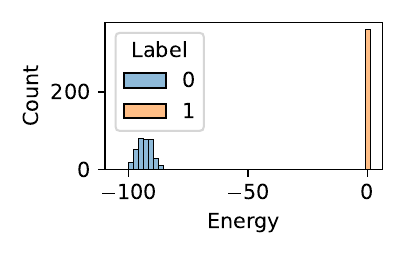}
            \caption{Open Q\&A}
            \label{fig:openqa_hist}
        \end{subfigure}
        &
        \begin{subfigure}[b]{0.24\textwidth}
            \centering
            \includegraphics[width=0.7\textwidth]{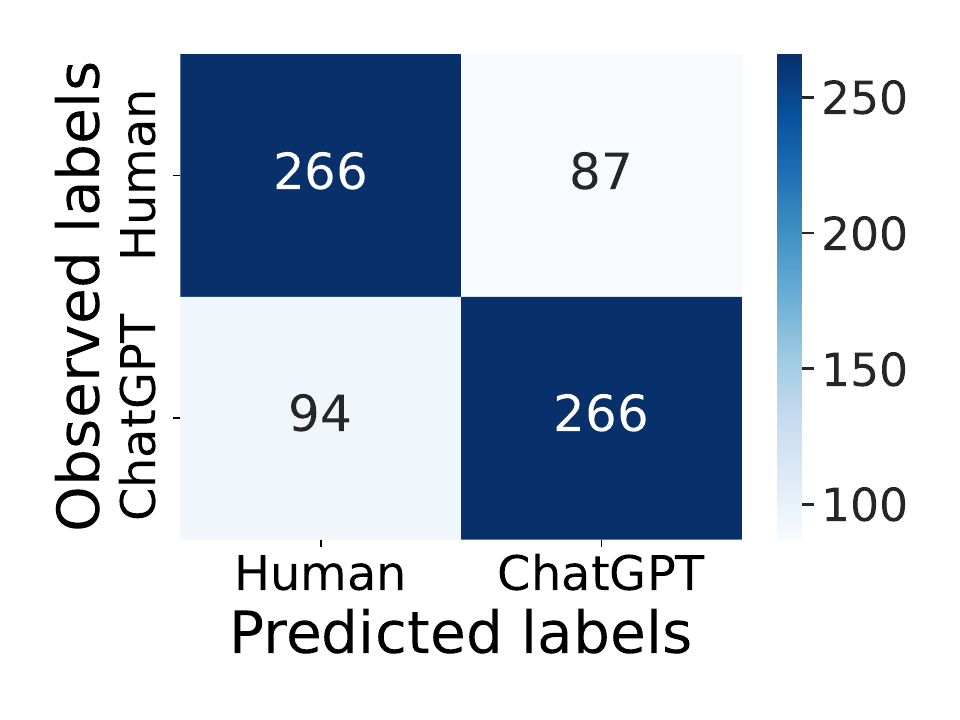}
            \caption{Open Q\&A}
            \label{fig:openqa_scatter}
        \end{subfigure}
        \\
        \begin{subfigure}[b]{0.24\textwidth}
            \centering
            \includegraphics[width=0.75\textwidth]{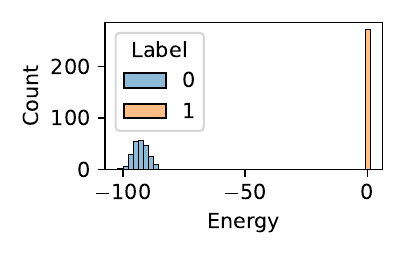}
            \caption{Wiki}
            \label{fig:wiki_hist}
        \end{subfigure}
        &
        \begin{subfigure}[b]{0.24\textwidth}
            \centering
            \includegraphics[width=0.7\textwidth]{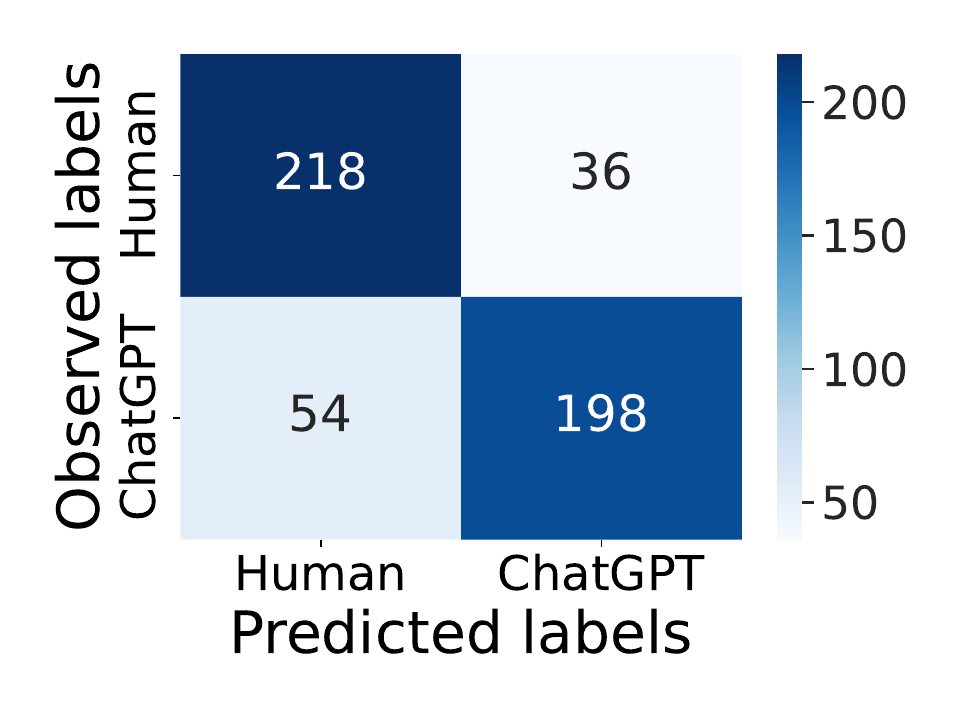}
            \caption{Wiki}
            \label{fig:wiki_scatter}
        \end{subfigure}
        &
        \begin{subfigure}[b]{0.24\textwidth}
            \centering
            \includegraphics[width=0.75\textwidth]{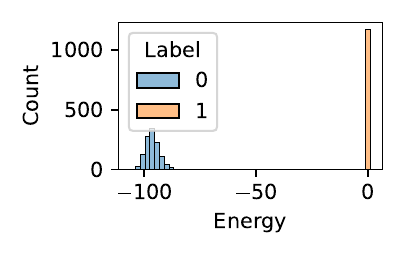}
            \caption{Finance}
            \label{fig:finance_hist}
        \end{subfigure}
        &
        \begin{subfigure}[b]{0.24\textwidth}
            \centering
            \includegraphics[width=0.7\textwidth]{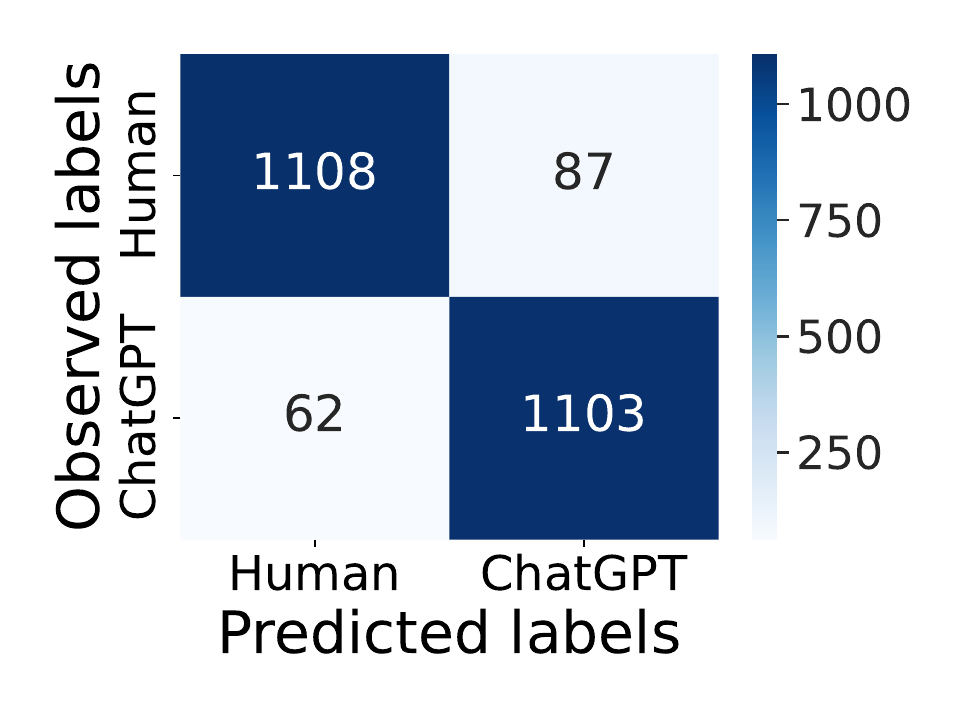}
            \caption{Finance}
            \label{fig:finance_scatter}
        \end{subfigure}
        \\
        \begin{subfigure}[b]{0.24\textwidth}
            \centering
            \includegraphics[width=0.75\textwidth]{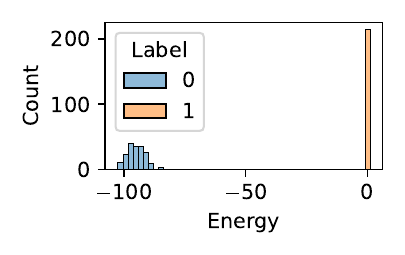}
            \caption{ArXiv}
            \label{fig:arxiv_hist}
        \end{subfigure}
        &
        \begin{subfigure}[b]{0.24\textwidth}
            \centering
            \includegraphics[width=0.7\textwidth]{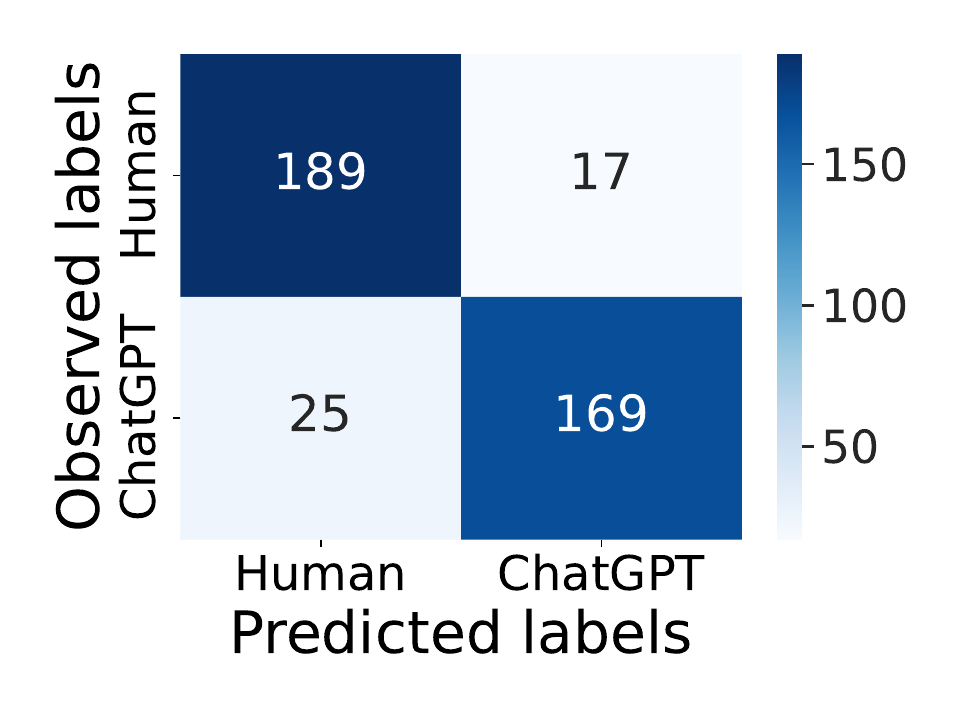}
            \caption{ArXiv}
            \label{fig:arxiv_scatter}
        \end{subfigure}
        &
        \begin{subfigure}[b]{0.24\textwidth}
            \centering
            \includegraphics[width=0.75\textwidth]{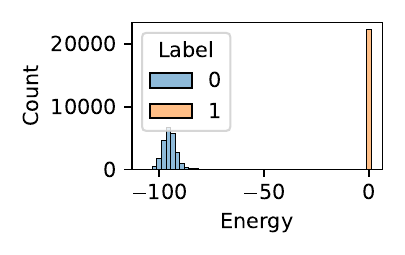}
            \caption{Combined}
            \label{fig:combined_hist}
        \end{subfigure}
        &
        \begin{subfigure}[b]{0.24\textwidth}
            \centering
            \includegraphics[width=0.7\textwidth]{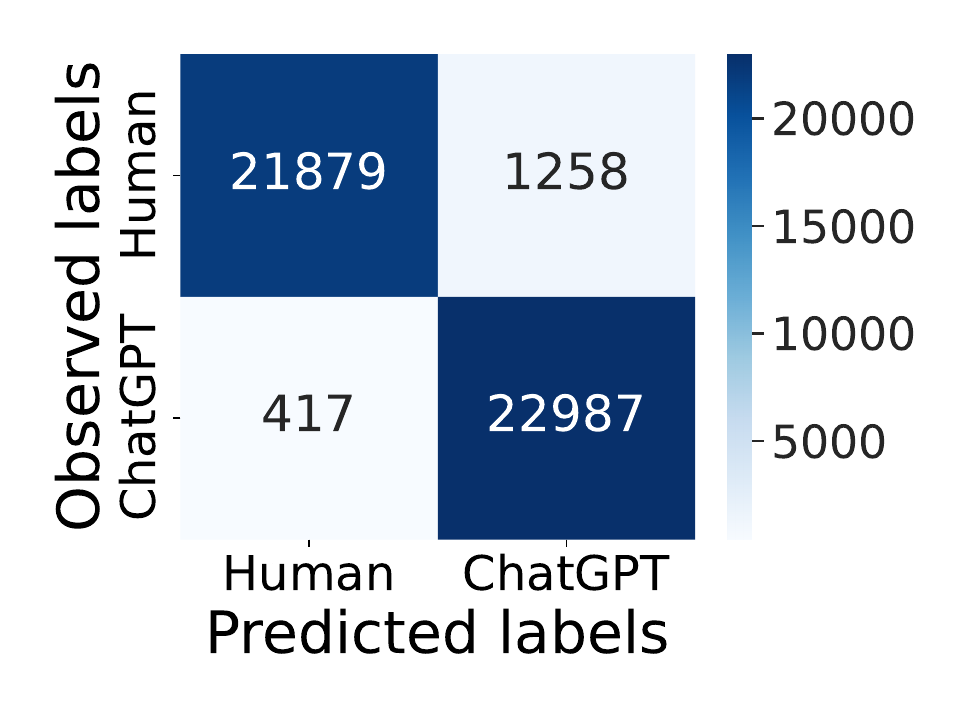}
            \caption{Combined}
            \label{fig:combined_scatter}
        \end{subfigure}
    \end{tabular}
    
    \caption{Different energy levels for different datasets. (a), (c), (e), (g), (i), (k), (m), and (o) represent the energy histogram plots for the Medical, Reddit, Political, Open Q\&A, Wiki, Finance, ArXiv, and the combined dataset. (b), (d), (f), (h), (j), (l), (n), and (p) represent the confusion matrix for the same.}
    \label{fig:diff_data_energy}
\end{figure*}

\begin{table}[t]
\centering
\small
\caption{Statistical results of sub-datasets in the experiments.}
\label{tab:subdatasetsenergy}

\begin{tabular}{|c|c|c|c|}
\hline
\multicolumn{1}{|c|}{\textbf{Dataset}} & \multicolumn{1}{c|}{\textbf{Records}} & \multicolumn{1}{c|}{\textbf{TPR} (\%)} & \multicolumn{1}{c|}{\textbf{TNR}(\%)} \\
\hline
Medical & 4,992 & 96.6 & 96.4 \\
\hline
Financial & 15,732 & 94.6 & 92.7 \\
\hline
Wiki & 3,368 & 78.5 & 85.8 \\
\hline
Open Q\&A & 4,748 & 73.9 & 75.4 \\
\hline
Reddit & 99,054 & 100.0 & 100.0 \\
\hline
Political & 3,620 & 95.2 & 87.7 \\
\hline
arXiv & 2,664 & 87.1 & 91.7 \\
\hline
Combined & 134,178 & 97.0 & 96.5 \\
\hline
\end{tabular}

\end{table}

\begin{table}[t]
\centering
\small
\caption{Statistical results of the \ourname in other domains.}
\label{tab:checkgpt_subdatasetsenergy}

\begin{tabular}{|c|c|c|c|}
\hline
\multicolumn{1}{|c|}{\textbf{Dataset}} & \multicolumn{1}{c|}{\textbf{Records}} & \multicolumn{1}{c|}{\textbf{TPR} (\%)} & \multicolumn{1}{c|}{\textbf{TNR}(\%)} \\
\hline
Task1 & 100000 & 94.7 & 97.9\\
\hline
Task2 & 100000 & 88.7 & 88.4 \\
\hline
Task3 & 100000 & 84.3 & 81.0 \\
\hline
\end{tabular}

\end{table}

\subsection{Energy distribution of human-written response and ChatGPT-generated content} 

We explore the differences in energy computation between human-generated responses and responses generated by ChatGPT. We visually represent the energy distribution for each dataset and highlight the significant variations in energy values between human and ChatGPT-generated answers. By examining the energy plots for each dataset, as depicted in Figure \ref{fig:diff_data_energy}, we can clearly observe distinct disparities in energy levels between human-generated text and ChatGPT-generated content. These plots specifically represent the perturbed energies of individual samples within the test set of each dataset. Figures \ref{fig:medical_hist}, \ref{fig:reddit_hist}, \ref{fig:political_hist}, \ref{fig:openqa_hist}, \ref{fig:wiki_hist}, \ref{fig:finance_hist}, \ref{fig:arxiv_hist}, and \ref{fig:combined_hist} illustrate the energy distribution for human-generated content and ChatGPT-generated content, showcasing two distinct levels. The lower values, which are highly negative, correspond to label 0 (ChatGPT-generated content), while the larger values belong to label 1 (human-generated content). Therefore, by analyzing the energy values, it becomes relatively easy to differentiate between human-generated and ChatGPT-generated content. Figures \ref{fig:medical_scatter}, \ref{fig:reddit_scatter}, \ref{fig:political_scatter}, \ref{fig:openqa_scatter}, \ref{fig:wiki_scatter}, \ref{fig:finance_scatter}, \ref{fig:arxiv_scatter}, and \ref{fig:combined_scatter} represents the energy distribution of human- and ChatGPT-generated content in relation to the EBM prediction.
It is important to highlight that misclassifications were observed for certain samples across different datasets, which introduces the possibility of reverse perturbed energies for both label options (0 or 1). Consequently,  confusion matrices for both label 0 and label 1, with distinct true positives (TP), true negatives (TN), false positives (FP), and false negatives (FN), are depicted in Figures \ref{fig:medical_scatter}, \ref{fig:political_scatter}, \ref{fig:openqa_scatter}, \ref{fig:wiki_scatter}, \ref{fig:finance_scatter}, \ref{fig:arxiv_scatter}, and \ref{fig:combined_scatter}. However, for the reddit dataset, the EBM achieved a testing accuracy of 100\%, as evidenced by Figure \ref{fig:reddit_hist}, where both true positive rate (TPR) and true negative rate (TNR) are equal to 100\%.

\begin{table}[t]
\centering
\small
\caption{Rephrasing statistical results for CheckGPT.}\label{tab:checkgpt}

\begin{tabular}{| c | c |c | c |} 
 \hline
\textbf{Dataset}  & $\#$\textbf{Records} & \textbf{TPR} (\%) & \textbf{TNR} (\%) \\  \hline
Task1   &   100,000    &    82.7  & 99.7         \\ \hline
Task2   &   100,000     &   56.3  & 96.1           \\ \hline
Task3   &   100,000     &   3.8  & 99.0         \\ \hline
\end{tabular}

\end{table}

\subsection{Rephrasing detection}
\label{sub:rephrase}
In the following, we evaluate the effectiveness of our approach, \ourname, by comparing it to the rephrasing of content generated by ChatGPT. As mentioned earlier, our benchmark dataset has been carefully curated to encompass various domains, and \ourname demonstrates remarkable performance on it. However, to evaluate the effectiveness of \ourname in a domain that differs from those covered in our dataset, we utilize the Task1, Task2, and Task3 datasets referenced in the research conducted by Liu et al. \cite{liu2023check}. We choose these datasets because the ChatGPT prompts predominantly begin with phrases like "In this work," "In this paper," or "This paper," which can introduce bias into the queries. Consequently, 
we will discuss that the detector proposed in Liu et al.'s work does not accurately function as demonstrated in their research, as we will showcase in the upcoming analysis. Therefore, in this study, we rephrase the abstract using these specific wordings and evaluate the EBM model designed for Task1, Task2, and Task3, respectively, as shown in Section \ref{sub:discipline}. \\
The effectiveness of CheckGPT \cite{liu2023check} was examined by testing it against the rephrased datasets of Task1, Task2, and Task3, and the results are summarized in Table \ref{tab:checkgpt}. The accuracy of CheckGPT decreased progressively as we evaluated it on the rephrased datasets. Specifically, for rephrased Task1, CheckGPT achieved an accuracy of 82.72\%, for rephrased Task2, it obtained an accuracy of 56.35\%, and for rephrased Task3, it achieved an accuracy of 3.76\%. These results indicate that CheckGPT does not deliver the expected performance when tasked with detecting the rephrased versions of academic abstracts. This discrepancy arises because CheckGPT was originally designed to identify academic abstracts in their original form, thus, introducing bias for the CheckGPT detection. \\ 
It is worth noting that the authors in \cite{liu2023check} developed distinct detectors for different tasks, training each task on separate models. In contrast, our approach involved testing the rephrased Task1, Task2, and Task3 against the EBM model trained on our combined benchmark dataset, to understand and measure how \ourname behaves when presented with never seen before samples. The accuracy achieved for rephrased Task1 was 76.9\%, for rephrased Task2 it was 68.7\%, and for rephrased Task3 it was 58.3\%, as shown in Table \ref{tab:rephrasing_combined}. However, when we trained separate EBMs for each original task and tested them on the rephrased versions, the true positive rates (TPR) obtained were 93\% for rephrased Task1, 83.4\% for rephrased Task2, and 75.4\% for rephrased Task3, as shown in Table \ref{tab:rephrasing_ours}. Thus, in comparison to CheckGPT's experimental evaluation results of 82.72\% for rephrased Task1, 56.35\% for rephrased Task2, and 3.76\% for rephrased Task3 (Table \ref{tab:checkgpt}), our proposed approach, \ourname, outperforms CheckGPT by a significant margin. 

Therefore, when we compare \ourname and \textit{CheckGPT}~\cite{liu2023check} under the same realistic settings, \ourname outperforms other detectors for the case of hybrid text, even if notably the accuracy is not as high as in our benchmark dataset.
 
\begin{table}[t]
\centering
\small
\caption{Statistical results for \ourname in other domains.}
\label{tab:rephrasing_combined}

\begin{tabular}{|c|c|c|c|}
\hline
\textbf{Dataset} & \textbf{\# Records} & \textbf{TNR} (\%) & \textbf{TPR} (\%) \\
\hline
Task1 & 100,000 & 76.9 & 76.6 \\
\hline
Task2 & 100,000 & 68.7 & 68.4 \\
\hline
Task3 & 100,000 & 58.3 & 58.9 \\
\hline
\end{tabular}

\end{table}

\begin{table}[t]
\centering
\small
\caption{Rephrasing statistical results for \ourname}\label{tab:rephrasing_ours}

\begin{tabular}{| c | c |c |c|} 
 \hline
\textbf{Dataset}  & $\#$\textbf{Records} & \textbf{TPR} (\%) & \textbf{TNR} (\%) \\  \hline
Task1   &   100,000    &    93.0  &  94.1         \\ \hline
Task2   &   100,000     &   83.4  & 84.5           \\ \hline
Task3   &   100,000     &   74.9  & 75.1         \\ \hline
\end{tabular}

\end{table}
\section{Related Works}
\label{sec:related}

\noindent
In general, automated machine learning techniques for identifying synthetic texts can be categorized into three primary groups: i) Basic classifiers \cite{guo2023close, solaiman2019release}, ii) Zero-shot detection approaches \cite{kushnareva2021artificial, mitchell2023detectgpt, zellers2019defending}, and iii) Detection methods based on fine-tuning \cite{mitrovic2023chatgpt}.

\noindent
\textbf{Basic classifiers.} The first category, simple classifiers, includes various detection techniques. For instance, OpenAI's logistic regression model \cite{solaiman2019release} was trained on TF-IDF, unigram, and bigram features and specifically analyzed text generated by \textit{GPT-2}. Through exploration of different generation strategies and model parameters, it was discovered that these classifiers can achieve accuracy levels of up to 97\%. However, detecting shorter outputs proves more challenging for these models compared to longer ones. Another simple classifier proposed by Guo et al. \cite{guo2023close} analyzed linguistic and stylistic characteristics of responses from ChatGPT and human experts. Unfortunately, these detection models were ineffective in identifying ChatGPT-generated text due to the imbalanced nature of the dataset they were trained on, failing to capture all of ChatGPT's text generation styles \cite{pegoraro2023chatgpt}. In a separate study, Kushnareva et al. \cite{kushnareva2021artificial} trained a logistic regression model using \textit{Topological Data Analysis} (TDA) to extract interpretable topological features, such as the number of connected components, edges, and cycles in the graph, for recognizing artificial text. However, this approach is unlikely to be effective for ChatGPT as it was not specifically tested on that particular model, but instead was conducted on datasets from WebText \& \textit{GPT-2}, Amazon Reviews \& \textit{GPT-2}, RealNews \& \textit{GROVER} (FakeNews)~\cite{zellers2019defending}. Furthermore, prior work \cite{pegoraro2023chatgpt} tested the authors \cite{kushnareva2021artificial} designed model and got an accuracy of 25.1\%.

\noindent
\textbf{Zero-shot detection.} Zero-shot detection techniques also includes a variety of approaches for identifying AI-generated text. For instance, Mitchell et al. \cite{mitchell2023detectgpt} introduced a method that leverages the log probabilities of the generative model to detect AI-generated text. However, this approach is primarily tailored to \textit{GPT-2} prompts and our evaluation indicates that it does not generalize well to \textit{GPT-3} models. In another study, the authors in \cite{zellers2019defending} utilized a transformer similar to the one employed in \textit{GPT-2} to create a tool called \textit{Grover}. \textit{Grover} is capable of generating text such as fake news and is also able to detect its own generated text. The study demonstrated the effectiveness of \textit{Grover} in verifying self-generated fake news. Nevertheless, it remains uncertain how well \textit{Grover} would perform on text generated by other \textit{GPT} models. Additionally, OpenAI \cite{solaiman2019release} developed a \textit{GPT-2} detector using a 1.5 billion parameter model that can accurately identify the top 40 generated outputs with an accuracy ranging from 83\% to 85\%. However, when fine-tuned on the Amazon reviews dataset, the accuracy dropped to 76\%.

\noindent
\textbf{Fine-tuned detection.} In case of fine-tuning based detection mechanism, Mitrović et al. \cite{mitrovic2023chatgpt} conducted a study exploring the training of a machine learning model that distinguishes ChatGPT-generated two-line restaurant reviews generated from real human-written reviews. They employed a framework based on \textit{DistilBERT}, which is a lightweight model based on \textit{BERT}. In another investigation, Solaiman et al. \cite{solaiman2019release} performed experiments to fine-tune pre-trained language models for the detection of AI-generated texts. They utilized \textit{$RoBERTa_{BASE}$} and \textit{$RoBERTa_{LARGE}$} as the basis for their classifiers. However, it is important to note that this approach failed to detect text generated by ChatGPT, as illustrated in \cite{rudolph2023chatgpt}. As a result, all the aforementioned detection methods are either exclusively applicable to prompts generated by the \textit{GPT-2} model or tailored for specific tasks.

\noindent
\textbf{Other methods.} Alternative methods exists outside the three main categories of detection approaches that deserve attention. These methods involve testing ChatGPT-generated text against plagiarism tools, developing Deep Neural Network-based detection tools, employing sampling-based approaches, and utilizing various online detection tools. 
Existing tools such as \textit{RoBERTa}, \textit{Grover}, and \textit{GPT-2} have been used to assess the originality of educational content compared to ChatGPT-generated text. For instance, a study evaluated the effectiveness of popular plagiarism-detection tools, \textit{iThenticate} and \textit{Turnitin}, in identifying plagiarism in 50 essays generated by ChatGPT \cite{khalil2023will}.
Another study by Gao et al. \cite{gao2022comparing} aimed to compare ChatGPT-generated academic paper abstracts using a \textit{GPT-2} Output Detector \textit{RoBERTa}, a plagiarism checker, and human review. The authors collected ten research abstracts from five high-impact medical journals and then used ChatGPT to output research abstracts based on their titles and journals. In a different study, Liu et al.\cite{liu2023check} developed a tool called CheckGPT, which is a content detector for LLMs  aimed at determining whether academic abstracts in the fields of computer science (CS), physics, humanities and social sciences (HSS) were generated by ChatGPT. The authors achieved a detection accuracy ranging from 98\% to 99\% for identifying ChatGPT-generated academic abstracts. Yet, as discussed in Section \ref{sec:eval}, when their CS Task1, Task2, and Task3 datasets underwent paraphrasing to eliminate potential biases stemming from phrases like "In this work," "In this paper," or "This paper," CheckGPT showed reduced true positive rates (TPR).

Nonetheless, prior research \cite{pegoraro2023chatgpt} has revealed the shortcomings of the discussed detection methods in accurately identifying text generated by ChatGPT. Additionally, in the same study, the authors evaluated multiple online tools \cite{originality, aicontentdetector, copyleaks, Huggingface, perplexityAnalysis, AITextClassifier, GPTZero, writefull, AITextDetector}, but the most effective tool for detecting generated text was found to have a success rate of less than 50\%. 

In contrast, we propose a detection tool that significantly distinguishes itself from previous methodologies by incorporating the characteristics of human speech and the thought process involved in communication with both humans and machines. Consequently, our approach outperforms other existing methods in detecting ChatGPT-generated text.
\section{Conclusion}
\label{sec:conclusion}
\noindent We propose the design and implementation of a novel detection tool named \ourname, which accurately determines the source of a given text by effectively differentiating between responses generated by the ChatGPT and those generated by humans. Unlike existing detection methods that focus solely on prompts generated by specific GPT models or serve specific purposes within certain fields (often yielding low accuracy), our approach is versatile and can successfully distinguish any text produced by other AI tools, especially ChatGPT, from human-generated text. By leveraging the intricate dynamics of human-to-human and human-to-machine communication, we incorporate the Doppler effect technique into our detection mechanism to account for inherent biases. Additionally, we address ChatGPT's rephrasing techniques by employing the explainable AI Integrated Gradients (IG) method. By integrating all these components, we construct an energy-based model that significantly enhances the detector's ability to accurately differentiate between responses generated by humans and ChatGPT.

We introduce a novel detection tool called \ourname, which accurately identifies the source of a given text by effectively distinguishing between responses generated by ChatGPT and those created by humans. Unlike existing detection methods, which are often limited to specific GPT models or specialized domains and consequently suffer from low accuracy, our approach is versatile. It can successfully discern any text, particularly from ChatGPT, against human-generated content.

Our approach leverages the intricate dynamics of human-to-human and human-to-machine communication, and incorporates the Doppler effect and energy-based model to our detection mechanism to account for inherent biases.  Furthermore, we address ChatGPT's rephrasing techniques by making use of the explainable AI method known as Integrated Gradients (IG).

\ourname significantly outperforms the existing ChatGPT detectors. Our approach goes beyond the limitations of existing methods, offering a more comprehensive and effective solution for source identification in the ever-evolving landscape of AI-generated text.

\section*{Acknowledgment}
This research received funding from Intel through the Private AI Collaborate Research Institute (\url{https://www.private-ai.org/}), and the Hessian Ministry of Interior and Sport as part of the F-LION project, following the funding guidelines for cyber security research.

\bibliographystyle{plain}
\bibliography{reference}

\appendices
\section*{Appendix}

\subsection{Large Language Models (LLMs)}
\label{sub:apx_llm}
\noindent Large Language Models represent sophisticated artificial intelligence systems specifically developed to comprehend and generate text that closely resembles human language. These models undergo extensive training on vast datasets, incorporating diverse sources such as books, articles, websites, and more. This training equips LLMs with the ability to generate coherent and contextually appropriate responses. Some notable examples of LLMs include: GPT-3 (Generative Pre-trained Transformer 3), GPT-2 (Generative Pre-trained Transformer 2), Transformer-XL, BERT (Bidirectional Encoder Representations from Transformers), and ChatGPT. The aforementioned models are merely a few examples of LLMs, as there exist numerous other variations and models crafted by different organizations and researchers. Each of these models possesses unique strengths and specializes in specific areas, contributing to the diverse landscape of LLM development.

\subsection{Drumhead vibrations}
\label{sub:apx_dv}
\noindent When the head of a drum is struck, it changes shape and compresses the air inside the shell. This compressed air then exerts pressure on the bottom head, causing it to change shape. These shape changes are transmitted to the drum shell, resulting in reflections and a repeated cycle that generates vibrations. The top and bottom head vibrations propagate into the surrounding air, becoming audible. As the vibrations of the drum heads gradually dampen, the sound diminishes.
The behavior of an idealized drumhead can be understood by examining the vibrations of a two-dimensional elastic membrane under tension \cite{sapoval1991vibrations} \cite{jenkins2006membrane}. This membrane, which can be modeled as a circular surface with uniform thickness connected to a rigid frame, exhibits fascinating properties. The membrane can store vibrational energy at specific resonant frequencies through the resonance phenomenon. The surface of the membrane moves in distinct patterns characterized by standing waves, referred to as normal modes. The lowest frequency normal mode is called the fundamental mode, and the membrane possesses an infinite number of these normal modes.

\end{document}